\documentclass[prd,preprint,showpacs]{revtex4}
\usepackage{amsmath}
\usepackage{bm}
\allowdisplaybreaks[1]
\usepackage{graphicx}

\newcommand{\diff}{\mathrm{d}}
\renewcommand{\vec}[1]{\bm{#1}}
\newcommand{\RPV}{\not\!{R}_p}
\newcommand{\modsq}[1]{\left|#1\right|^2}
\newcommand{\lambsq}{\lambda'^2}
\newcommand{\lambsqd}{\lambsq\mu_r^{2\epsilon}}
\DeclareMathOperator{\acosh}{arccosh}

\begin{document}

\title{Soft Gluon Resummation Effects in Single Slepton Production at Hadron Colliders}
\author{Li Lin Yang}
\email{llyang@pku.edu.cn}
\author{Chong Sheng Li}
\email{csli@pku.edu.cn}
\author{Jian Jun Liu}
\altaffiliation{Present address: Department of Physics, Tsinghua University, Beijing 100084, China}
\author{Qiang Li}
\affiliation{Department of Physics, Peking University, Beijing 100871, China}
\date{\today}

\begin{abstract}
  We investigate QCD effects in the production of a single slepton at hadron colliders in
  the Minimal Supersymmetric Standard Model without R-parity.  We calculate the total
  cross sections and the transverse momentum distributions at next-to-leading order in
  QCD. The NLO corrections enhance the total cross sections and decrease the dependence
  of the total cross sections on the factorization and renormalization scales. For the
  differential cross sections, we resum all order soft gluon effects to give reliable
  predictions for the transverse momentum distributions. We also compare two approaches
  to the non-perturbative parametrization and found that the results are slightly
  different at the Tevatron and are in good agreement at the LHC. Our results can be
  useful to the simulation of the events and to the future collider experiments.
\end{abstract}


\maketitle

\section{Introduction}

The Minimal Supersymmetric Standard Model (MSSM) is one of the most popular new physics
models beyond the Standard Model (SM). In the MSSM, R-parity conservation is imposed in
order to keep the proton stable. R-parity is a discrete symmetry, which is defined to be
$R_p \equiv (-1)^{3(B-L)+2S}$, where $B$, $L$ and $S$ are baryon number, lepton number,
and spin, respectively. All SM particles have $R_p = 1$, and all superpartners have $R_p
= -1$.  In consequence, the superpartners can only be produced in pair, and the lightest
superpartner (LSP) is stable. However, proton decay may be avoided by assuming $B$ or $L$
conservation, but not both. In these cases, R-parity can be violated. For a recent review
of the R-parity violating ($\RPV$) extension of the MSSM, see Ref.~\cite{hep-ph/0406039}.

The most general form for the $\RPV$ part of the superpotential is
\begin{equation}\label{eq:sp}
  W_{\RPV} = \mu_iL_iH_2 + \frac{1}{2}\lambda_{ijk}L_iL_jE_k^c +
  \lambda'_{ijk}L_iQ_jD_k^c + \frac{1}{2}\lambda''_{ijk}U_i^cD_j^cD_k^c,
\end{equation}
where $L_i(Q_i)$ and $E_i(U_i,D_i)$ are, respectively, the left-handed lepton (quark)
$SU(2)$ doublet and right-handed lepton (quark) $SU(2)$ singlet chiral superfields, and
$H_{1,2}$ are the Higgs chiral superfields. The indices $i,j,k$ denote generations and
the superscript $c$ denotes charge conjugation. As stated before, we require that the
lepton number violating part and the baryon number violating part should not exist
simultaneously. For the purpose of this paper, we will assume that only the lepton number
violating couplings $\lambda'_{ijk}$ are non-zero. A recent summary of the bounds on the
couplings in Eq.~(\ref{eq:sp}) can be found in Ref.~\cite{hep-ph/0406029}.

The $\RPV$-MSSM has the remarkable feature that sparticles need not be produced in pair.
Therefore, a definitive signal of R-parity violation will be the observation of single
sparticle production. The terms in Eq.~(\ref{eq:sp}) will result in various resonant
sparticle production processes. The term involving $\lambda_{ijk}$ leads to resonant
sneutrino production at $e^+e^-$ colliders \cite{epem}, while the $\lambda''_{ijk}$ term
leads to resonant squark production in hadron-hadron collisions
\cite{Phys.Rev.D41.2099,sqhh}. The term with $\lambda'_{ijk}$ can induce both resonant
squark production at $ep$ colliders \cite{Nucl.Phys.B397.3} and resonant slepton
production at hadron-hadron colliders \cite{Phys.Rev.D41.2099,slhh,Phys.Rev.D63.055008}.
We will consider the last case in the following.

The resonant production of a single slepton can lead to interesting phenomenology at
hadron colliders. A charged slepton can decay into a neutralino and a charged lepton, and
the neutralino can subsequently decay into a charged lepton and two jets via $\lambda'$
couplings.  Due to the Majorana nature of the neutralino, the two leptons can have either
opposite or same charges. The case of two leptons of the same charges is more interesting
due to the absence of large SM background. For a detailed analysis of the signal and the
background, see Ref.~\cite{Phys.Rev.D63.055008}. For a sneutrino, the decay products can
be a chargino and a charged lepton. With the subsequent decay of the chargino, the final
state could contain three charged leptons which is also a clean signature.

The analysis in the above is all based on the tree-level cross sections. It is well known
that in order to reduce the scale dependence of the cross section and get more precision
predictions, one must include the QCD radiative corrections. The next-to-leading order
QCD corrections to the total cross sections have been calculated in
Ref.~\cite{Nucl.Phys.B660.343}.  However, they used rather old parton distribution
functions in the numerical calculations, and their results have large dependence on the
PDF sets used. Moreover, they did not consider the kinematic distribution of the events,
which is very important in designing the strategy of discovery. In this paper, we will
calculate the transverse momentum distribution of the slepton at next-to-leading order in
QCD, and also calculate all order soft gluon resummation effects to give reasonable
predictions. In our numerical evaluations, we will use the new sets of PDFs, and compare
our predictions for the total cross sections with those of
Ref.~\cite{Nucl.Phys.B660.343}.

The paper is organized as follows. Section~\ref{sec:proc} describes the production
processes at hadron/parton level and shows the leading order (LO) cross sections.
Section~\ref{sec:nlo} presents the analytical expression of the next-to-leading order
(NLO) corrections to the cross sections. Section~\ref{sec:total} gives the analytical and
numerical results for the total cross sections and Section~\ref{sec:qt} gives the
transverse momentum distribution. And finally is the summary.

\section{The processes and the cross sections}\label{sec:proc}

We consider the process $A(p_1) + B(p_2) \to L(q) + X$, where $A$ and $B$ are the
incoming hadrons (proton-antiproton for Tevatron or proton-proton for LHC) with momenta
$p_1$ and $p_2$, $L$ stands for charged slepton $\tilde{l}$ (or sneutrino $\tilde{\nu}$)
with momentum $q$. At parton level, four kinds of subprocesses can be induced by the
$\lambda'_{ijk}$ couplings at tree level: $d_k + \bar{u}_j \to \tilde{l}_i$, $d_j +
\bar{d}_k \to \tilde{\nu}_i$, and their charge conjugated processes.

In the QCD improved parton model, the hadronic differential cross section can be
factorized into the convolution of the partonic differential cross sections with
appropriate parton distribution functions (PDFs):
\begin{equation}
  \frac{\diff\sigma^{AB}}{\diff{q_T^2}\diff{y}} = \sum_{\alpha,\beta} \int \diff{x_1}
  \diff{x_2} \frac{\diff\hat{\sigma}^{\alpha\beta}}{\diff{q_T^2}\diff{y}}
  f_{\alpha/A}(x_1,\mu_f) f_{\beta/B}(x_2,\mu_f),
\end{equation}
where $\hat{\sigma}^{\alpha\beta}$ is the cross section for the partonic subprocess
$\alpha(\hat{p}_1) + \beta(\hat{p}_2) \to L(q) + X$. $\hat{p}_1$ and $\hat{p}_2$ are the
momenta of the incoming partons $\alpha$, $\beta$. The momenta fractions $x_1$ and $x_2$
are defined by $x_i=\hat{p}_i/p_i, (i = 1, 2)$. $f_{p/H}(x,\mu_f)$ is the parton
distribution function which describes the probability of finding a parton $p$ with
momentum fraction $x$ inside the hadron $H$ at factorization scale $\mu_f$. The sum is
over all possible initial partons which contribute. For any momentum $q$, it can be
decomposed as
\begin{equation}
  q = (q^+, q^-, \vec{q}_T), \qquad q_T^2 = \vec{q}^2_T,
\end{equation}
where $q^\pm = q^0 \pm q^3$, and the rapidity $y$ is defined as
\begin{equation}
  y = \frac{1}{2}\ln\frac{q^+}{q^-}.
\end{equation}
We work in the center-of-mass frame of the colliding hadrons, in which
\begin{equation}
  p_1 = (\sqrt{s}, 0, \vec{0}_T), \qquad p_2 = (0, \sqrt{s}, \vec{0}_T),
\end{equation}
where $s=(p_1+p_2)^2$ is the center-of-mass energy squared.

The partonic cross section $\hat{\sigma}^{\alpha\beta}$ can be calculated order by order
in perturbation theory. The leading order contribution is the Born level process
$\alpha(\hat{p}_1) + \beta(\hat{p}_2) \to L(q)$. Here $\alpha$ and $\beta$ represent only
quarks and anti-quarks. The matrix element is simply
\begin{equation}
  \mathcal{M}_B = \lambda'_{ijk} \bar{v}(\hat{p}_2) P_L u(\hat{p}_1)
\end{equation}
where $P_L=(1-\gamma^5)/2$, and we have ignored the mixing between the left-handed and
right-handed sleptons since they are almost degenerate. The coupling $\lambda'_{ijk}$
depends on the generations of $\alpha$, $\beta$ and $L$. For simplicity, we will make the
subscript implicit below and simply write $\lambda'$. The one particle phase space can be
written as
\begin{align}
  \diff\Phi_1 &= \frac{\diff^3q}{(2\pi)^32q_0} (2\pi)^4 \delta^4(\hat{p}_1+\hat{p}_2-q)
  \notag \\
  &= (2\pi) \diff{q_T^2} \diff{y} \, \delta(q_T^2) \delta(\hat{p}_1^+-q^+)
  \delta(\hat{p}_2^--q^-).
\end{align}
After convoluted with the PDFs, the differential cross section with respect to the
transverse momentum and rapidity of the slepton is
\begin{align}
  \frac{\diff\sigma_B}{\diff{q_T^2}\diff{y}} &= \frac{\pi}{m^2s}
  \overline{\modsq{\mathcal{M}_B}} \, \delta(q_T^2) f_{\alpha/A}(x_1^0,\mu_f)
  f_{\beta/B}(x_2^0,\mu_f) \notag
  \\
  &= \frac{\pi}{12s} \lambda'^2 \, \delta(q_T^2) f_{\alpha/A}(x_1^0,\mu_f)
  f_{\beta/B}(x_2^0,\mu_f),
\end{align}
where $x_1^0=\sqrt{\tau}e^y$ and $x_2^0=\sqrt{\tau}e^{-y}$ with $\tau=m^2/s$, and $m$ is
the mass of the slepton. We have made the summation over the initial states $\alpha$,
$\beta$ implicit.

\section{Next-to-leading order calculations} \label{sec:nlo}

The NLO QCD corrections consist of the following contributions: the exchange of virtual
gluons and the corresponding renormalization counterterms, the real gluon emission
subprocesses, the gluon initiated subprocesses, and the contributions of Altarelli-Parisi
(A-P) splitting functions. In the following, we will calculate these contributions
seperately. We use dimensional regularization (DREG) in $d=4-2\epsilon$ dimensions to
regulate all divergences and adopt $\overline{\text{MS}}$ renomalization and factorization
scheme to remove the ultraviolet (UV) and infrared (IR) (including soft and collinear)
divergences.

\subsection{Virtual gluon exchange}

Evaluating the relevant Feynman diagrams of virtual corrections, we obtain the amplitude
\begin{equation}
  \mathcal{M}_V = \mathcal{M}_B \frac{\alpha_s}{2\pi} C_F (4\pi)^\epsilon
  \Gamma(1+\epsilon) \left[ -\frac{1}{\epsilon^2} - \frac{1}{\epsilon}
    \ln\frac{\mu_r^2}{m^2} - \frac{1}{2} \ln^2\frac{\mu_r^2}{m^2} + \frac{2\pi^2}{3} - 1
  \right],
\end{equation}
where $C_F = 4/3$ and $\mu_r$ is the renormalization scale. In order to remove the UV
divergences in $\mathcal{M}_V$, a renormalization procedure must be carried out. We
define the renormalization constants as
\begin{align}
  \lambda'_0 &= Z_\lambda \lambda' \mu_r^\epsilon= (1 + \delta Z_\lambda) \lambda'
  \mu_r^\epsilon, \\
  \psi_0 &= Z_{\psi{L}}^{1/2} \psi_L + Z_{\psi{R}}^{1/2} \psi_R \notag \\
  &= \left( 1 + \frac{1}{2} \delta Z_{\psi{L}} \right) \psi_L + \left( 1 + \frac{1}{2}
    \delta Z_{\psi{R}} \right) \psi_R,
\end{align}
where $\lambda'_0$ and $\psi_0$ are the bare coupling and the bare quark wave function,
respectively. In the $\overline{\text{MS}}$ scheme, these renormalization constants are
fixed to be
\begin{align}
  \delta Z_{\psi{L}} &= \delta Z_{\psi{R}} = - \frac{\alpha_s}{4\pi} C_F (4\pi)^\epsilon
  \Gamma(1+\epsilon) \frac{1}{\epsilon} , \\
  \delta Z_\lambda &= - \frac{\alpha_s}{4\pi} C_F (4\pi)^\epsilon \Gamma(1+\epsilon)
  \frac{3}{\epsilon}.
\end{align}
Thus the running of the coupling constant is governed by
\begin{equation}
  \lambda'(\mu) = \frac{\lambda'(\mu_0)}{1 + \displaystyle{\frac{3\alpha_s}{4\pi} C_F
      \log{\frac{\mu^2}{\mu_0^2}}}}.
\end{equation}

After adding the counter term, the UV divergences in $\mathcal{M}_V$ are cancelled, but
the IR divergent terms still persist. The corresponding differential cross section is
\begin{equation}
  \frac{\diff\sigma_V}{\diff{q_T^2}\diff{y}} = \frac{\diff\sigma_B}{\diff{q_T^2}\diff{y}}
  \frac{\alpha_s}{\pi} C_F (4\pi)^\epsilon \Gamma(1+\epsilon) \left[
    -\frac{1}{\epsilon^2} - \frac{1}{\epsilon} \left( \frac{3}{2} +
      \ln\frac{\mu_r^2}{m^2} \right) - \frac{1}{2} \ln^2\frac{\mu_r^2}{m^2} +
    \frac{2\pi^2}{3} - 1 \right].
\end{equation}

\subsection{Real gluon emission}

We now consider the contribution from the emission of one gluon $\alpha(\hat{p}_1) +
\beta(\hat{p}_2) \to L(q) + g(k)$. We define the Mandelstam variables as
\begin{align}
  s &= (p_1+p_2)^2, \notag
  \\
  t &= (p_1-q)^2 = m^2 - p_1^+q^- = m^2 - \sqrt{s(q_T^2+m^2)}e^{-y}, \notag
  \\
  u &= (p_2-q)^2 = m^2 - p_2^-q^+ = m^2 - \sqrt{s(q_T^2+m^2)}e^y, \notag
  \\
  \hat{s} &= (\hat{p}_1+\hat{p}_2)^2 = x_1x_2s, \notag
  \\
  \hat{t} &= (\hat{p}_1-q)^2 = x_1(t-m^2)+m^2, \notag
  \\
  \hat{u} &= (\hat{p}_2-q)^2 = x_2(u-m^2)+m^2.
\end{align}
The squared matrix element (with spin and color summed and averaged) can be expressed as
\begin{align}
  \overline{\modsq{\mathcal{M}_R}} &= \frac{8}{9} \pi\alpha_s \lambsqd
  \frac{(1-\epsilon)(\hat{s}^2+m^4)+2\epsilon\hat{s}m^2}{\hat{t}\hat{u}} \notag
  \\
  &= \frac{8}{9} \pi\alpha_s \lambsqd x_1x_2s \frac{1}{q_T^2} \left\{
    (1-\epsilon) \left[ 1 + \left(\frac{\tau}{x_1x_2}\right)^2 \right] + 2\epsilon
    \frac{\tau}{x_1x_2} \right\}.
\end{align}
The two particle phase space in $d$ dimensions is
\begin{align}
  \diff\Phi_2 &= \frac{\diff^{d-1}q}{(2\pi)^{d-1}2q_0}
  \frac{\diff^{d-1}k}{(2\pi)^{d-1}2k_0} (2\pi)^d \delta^d(\hat{p}_1+\hat{p}_2-q-k) \notag
  \\
  &= \frac{1}{2(2\pi)^{d-2}} \diff^{d-2}\vec{q}_T \diff{y} \,
  \delta(\hat{s}+\hat{t}+\hat{u}-m^2) \notag
  \\
  &= \frac{1}{8\pi} \frac{1}{\Gamma(1-\epsilon)} \left(\frac{4\pi}{q_T^2}\right)^\epsilon
  \diff{q_T^2} \diff{y} \, \delta(\hat{s}+\hat{t}+\hat{u}-m^2).
\end{align}
Thus the differential cross section can be written as
\begin{align}
  \frac{\diff\sigma_R}{\diff{q_T^2}\diff{y}} &= \int \diff{x_1} \diff{x_2} F(x_1,x_2)
  \delta(x_1x_2s+x_1(t-m^2)+x_2(u-m^2)+m^2) \notag
  \\
  &= \frac{1}{s} \int_{x_1^-}^1 \frac{\diff{x_1}}{x_1-x_1^+} F(x_1,x_2^*),
\end{align}
where
\begin{align}
  x_1^- &= \frac{x_1^+-\tau}{1-x_2^+}, \qquad x_2^*=\frac{x_1x_2^+-\tau}{x_1-x_1^+},
  \notag \\
  x_1^+ &= e^{y}\sqrt{\tau+q_T^2/s}, \qquad x_2^+=e^{-y}\sqrt{\tau+q_T^2/s},
\end{align}
and
\begin{align}
  F(x_1,x_2) &= \frac{\alpha_s}{18} \lambsq \frac{1}{\Gamma(1-\epsilon)}
  \left(\frac{4\pi\mu_r^2}{q_T^2}\right)^\epsilon f_{\alpha/A}(x_1,\mu_f)
  f_{\beta/B}(x_2,\mu_f)
  \notag \\
  &\qquad \times \frac{1}{q_T^2} \left\{ (1-\epsilon) \left[ 1 +
      \left(\frac{\tau}{x_1x_2}\right)^2 \right] + 2\epsilon \frac{\tau}{x_1x_2}
  \right\}.
\end{align}

\subsection{Gluon splitting subprocesses}

In addition to the real gluon emission subprocess, there are also contributions from the
gluon initiated processes $\alpha(\hat{p}_1) + g(\hat{p}_2) \to L(q) + \bar{\beta}(k)$
and $g(\hat{p}_1) + \beta(\hat{p}_2) \to L(q) + \bar{\alpha}(k)$. Defining the Mandelstam
variables as before, for the first subprocesses we can write the the squared matrix
elements as
\begin{align}
  \overline{\modsq{\mathcal{M}^{qg}_G}} &= \frac{1}{3} \pi\alpha_s \lambsqd
  \frac{(1-\epsilon)(\hat{u}^2+m^4)+2\epsilon\hat{u}m^2}{-(1-\epsilon)\hat{s}\hat{t}}
  \notag \\
  &= \frac{1}{3} \pi\alpha_s \lambsqd x_1x_2s \frac{1}{q_T^2} \left\{
    \frac{(x_2x_1^+-\tau)[(x_2x_1^+-\tau)^2+\tau^2]}{(x_1x_2)^3} -
    \frac{2\epsilon}{1-\epsilon} \frac{(x_2x_1^+-\tau)^2\tau}{(x_1x_2)^3} \right\},
\end{align}
while the squared matrix elements for the second subprocess can be obtained from the
above one by the substitution $\hat{t} \leftrightarrow \hat{u}$, $x_1 \leftrightarrow
x_2$, $x_1^+ \leftrightarrow x_2^+$, namely
\begin{align}
  \overline{\modsq{\mathcal{M}^{gq}_G}} &= \frac{1}{3} \pi\alpha_s \lambsqd
  \frac{(1-\epsilon)(\hat{t}^2+m^4)+2\epsilon\hat{t}m^2}{-(1-\epsilon)\hat{s}\hat{u}}
  \notag \\
  &= \frac{1}{3} \pi\alpha_s \lambsqd x_1x_2s \frac{1}{q_T^2} \left\{
    \frac{(x_1x_2^+-\tau)[(x_1x_2^+-\tau)^2+\tau^2]}{(x_1x_2)^3} -
    \frac{2\epsilon}{1-\epsilon} \frac{(x_1x_2^+-\tau)^2\tau}{(x_1x_2)^3} \right\}.
\end{align}
Thus the differential cross section is
\begin{align}
  \frac{\diff\sigma_G}{\diff{q_T^2}\diff{y}} &= \int \diff{x_1} \diff{x_2} F'(x_1,x_2)
  \delta(x_1x_2s+x_1(t-m^2)+x_2(u-m^2)+m^2) \notag \\
  &= \frac{1}{s} \int_{x_1^-}^1 \frac{\diff{x_1}}{x_1-x_1^+} F'(x_1,x_2^*),
\end{align}
where
\begin{align}
  F'(x_1,x_2) &= \frac{1}{16\pi} \frac{1}{x_1x_2s} \frac{1}{\Gamma(1-\epsilon)}
  \left(\frac{4\pi}{q_T^2}\right)^\epsilon
  \notag \\
  &\qquad \times \left[ f_{\alpha/A}(x_1,\mu_f) f_{g/B}(x_2,\mu_f)
    \overline{\modsq{\mathcal{M}^{qg}_G}} + f_{g/A}(x_1,\mu_f) f_{\beta/B}(x_2,\mu_f)
    \overline{\modsq{\mathcal{M}^{gq}_G}} \right].
\end{align}

\subsection{Splitting function contributions}

The contributions from splitting functions are essential to cancel the collinear
divergences in the cross sections. They arise from the renormalization of the PDFs at
NLO.  In the $\overline{\text{MS}}$ scheme, the renormalized PDFs can be expressed as
\begin{equation}\label{eq:pdf}
  f_{\alpha/H}(x,\mu_f) = f_{\alpha/H}(x) + \sum_{\beta} \left( -\frac{1}{\epsilon}
  \right) \frac{\alpha_s}{2\pi} \Gamma(1+\epsilon)
  \left(\frac{4\pi\mu_r^2}{\mu_f^2}\right)^\epsilon \int_x^1 \frac{\diff{z}}{z}
  P_{\alpha\beta}(z) f_{\beta/H}(x/z),
\end{equation}
where $P_{\alpha\beta}(z)$ are the A-P splitting functions, which are given by
\begin{align}
  P_{qq}(z) &= P_{\bar{q}\bar{q}}(z) = C_F \left[ \frac{3}{2}\delta(1-z) +
    \frac{1+z^2}{(1-z)_+} \right], \\
  P_{qg}(z) &= P_{\bar{q}g}(z) = \frac{1}{2} (z^2 + (1-z)^2).
\end{align}

The resulting contributions to the differential cross section are
\begin{align}
  \frac{\diff\sigma_C}{\diff{q_T^2}\diff{y}} &= \frac{\pi}{12s} \lambsq \delta(q_T^2)
  \frac{1}{\epsilon} \frac{\alpha_s}{2\pi} \Gamma(1+\epsilon)
  \left(\frac{4\pi\mu_r^2}{\mu_f^2}\right)^\epsilon
  \notag \\
  &\qquad \times \left[ (P\circ{f})_{\alpha/A}(x_1^0,\mu_f) f_{\beta/B}(x_2^0,\mu_f) +
    f_{\alpha/A}(x_1^0,\mu_f) (P\circ{f})_{\beta/B}(x_2^0,\mu_f) \right],
\end{align}
where
\begin{equation}
  (P\circ{f})_{\alpha/H}(x,\mu_f) = \sum_{\gamma} \int_{x}^1 \frac{\diff{z}}{z}
    P_{\alpha\gamma}(z) f_{\gamma/H}(x/z,\mu_f).
\end{equation}

\section{Total cross sections}\label{sec:total}

The total cross sections can be obtained by integrating out $q_T$ and $y$ from the
differential cross sections given above. The leading order result is
\begin{equation}
  \sigma_{\text{LO}} = \frac{\pi}{12s} \lambsq \int_{\tau}^1 \frac{\diff x}{x}
  f_{\alpha/A}(x,\mu_f) f_{\beta/B}(\tau/x,\mu_f).
\end{equation}
The next-to-leading order result can be written as
\begin{align}
  \sigma_{\text{NLO}} &= \int \diff{x_1} \diff{x_2} \biggl[ \hat{\sigma}^{\alpha\beta}
    f_{\alpha/A}(x_1,\mu_f) f_{\beta/B}(x_2,\mu_f) \biggr. \notag
  \\
  &\qquad + \biggl. \hat{\sigma}^{\alpha{g}} f_{\alpha/A}(x_1,\mu_f) f_{g/B}(x_2,\mu_f) +
    \hat{\sigma}^{g\beta} f_{g/A}(x_1,\mu_f) f_{\beta/B}(x_2,\mu_f) \biggr],
\end{align}
where
\begin{align}
  \hat{\sigma}^{\alpha\beta} &= \frac{\alpha_s}{9\hat{s}} \lambsq \left\{ \left(
      \frac{3}{2}\ln\frac{\mu_r^2}{\mu_f^2} + \frac{\pi^2}{3} - 1 \right) \delta(1-z) +
    \ln\frac{m^2}{\mu_f^2}\frac{1+z^2}{(1-z)_+} \right.
  \notag \\
  &\quad + \left. 2(1+z^2)\left(\frac{\ln(1-z)}{1-z}\right)_+ - (1+z^2)\frac{\ln{z}}{1-z}
    + (1-z) \right\},
  \\
  \hat{\sigma}^{\alpha{g}} &= \hat{\sigma}^{g\beta} = \frac{\alpha_s}{48\hat{s}}
  \lambsq \left[ \left( \ln\frac{m^2}{\mu_f^2} + \ln\frac{(1-z)^2}{z} \right)
    (z^2+(1-z)^2) + \frac{1}{2}(1-z)(7z-3) \right].
\end{align}
Here $\displaystyle{z \equiv \frac{m^2}{\hat{s}} = \frac{\tau}{x_1x_2}}$, and the
function with a subscript ``$+$'' is a distribution, which is defined as
\begin{equation}
  \int_0^1 \diff z g(z) f(z)_+ = \int_0^1 \diff z (g(z)-g(1)) f(z)
\end{equation}
for an arbitrary function $g(z)$. Our expressions for the total cross sections are the
same as those obtained in Ref.~\cite{Nucl.Phys.B660.343}.

In the numerical evaluation of the total cross sections, we use the updated version of
the CTEQ and MRST PDF, namely CTEQ6.1 and MRST2004 (there is no update for GRV98 PDF).
The renormalization and factorization scale are taken to the mass of the final state
slepton, $\mu_r=\mu_f=m$, unless otherwise specified. We will set the initial quarks to
only the light flavours $u$, $d$ and $s$. Due to the severe constraints on the products
of two $\lambda'$s, we will consider the contributions from one single $\lambda'_{ijk}$
at a time. The most stringent constraints on the couplings relevant here come from the
rare semileptonic decay $K \to \pi \nu \bar{\nu}$, which gives $\lambda'_{ijk} \leq
0.012m_{\tilde{d}_k}/(100\text{GeV})$ for $k=1,2$. We can assume that the first two
generations of sfermions have nearly degenerate masses, so we can take the down squark
mass as the slepton mass approximately. In the following, unless otherwise specified, we
will take the tree level coupling $\lambda'_{ijk}=0.01$, which can satisfy the
constraints for all the slepton masses under consideration. The cross sections for other
values of $\lambda'$ can be obtained easily by a multiplicative factor.

For convenience, we define three cross sections as following:
\begin{align}
  \sigma_{\mathrm{LO1}}: &\quad \text{LO partonic cross section convoluted with LO PDFs};
  \notag \\
  \sigma_{\mathrm{LO2}}: &\quad \text{LO partonic cross section convoluted with NLO
    ($\overline{\text{MS}}$) PDFs}; \notag
  \\
  \sigma_{\mathrm{NLO}}: &\quad \text{NLO partonic cross section convoluted with NLO
    ($\overline{\text{MS}}$) PDFs}, \notag
\end{align}
and correspondingly two $K$ factors:
\begin{equation*}
  K_1 = \frac{\sigma_{\mathrm{NLO}}}{\sigma_{\mathrm{LO1}}}, \qquad
  K_2 = \frac{\sigma_{\mathrm{NLO}}}{\sigma_{\mathrm{LO2}}}.
\end{equation*}
As the above definitions, $K_2$ measures only the size of the NLO QCD corrections to the
cross sections, while $K_1$ accounts for the effects of changing parton distribution
functions additionally.

\subsection{Sneutrino production}

We will consider sneutrino production first. The possible partonic initial states are
$d\bar{d}$, $d\bar{s}$, $s\bar{d}$ and $s\bar{s}$.

\begin{figure}[ht!]
  \centering
  \includegraphics{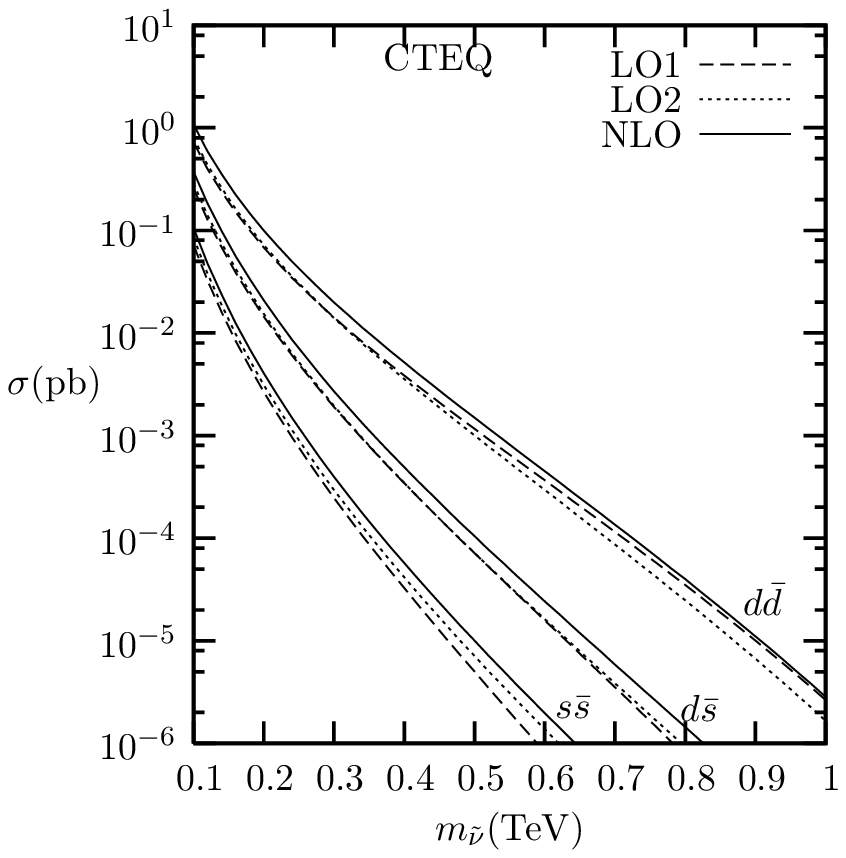}
  \includegraphics{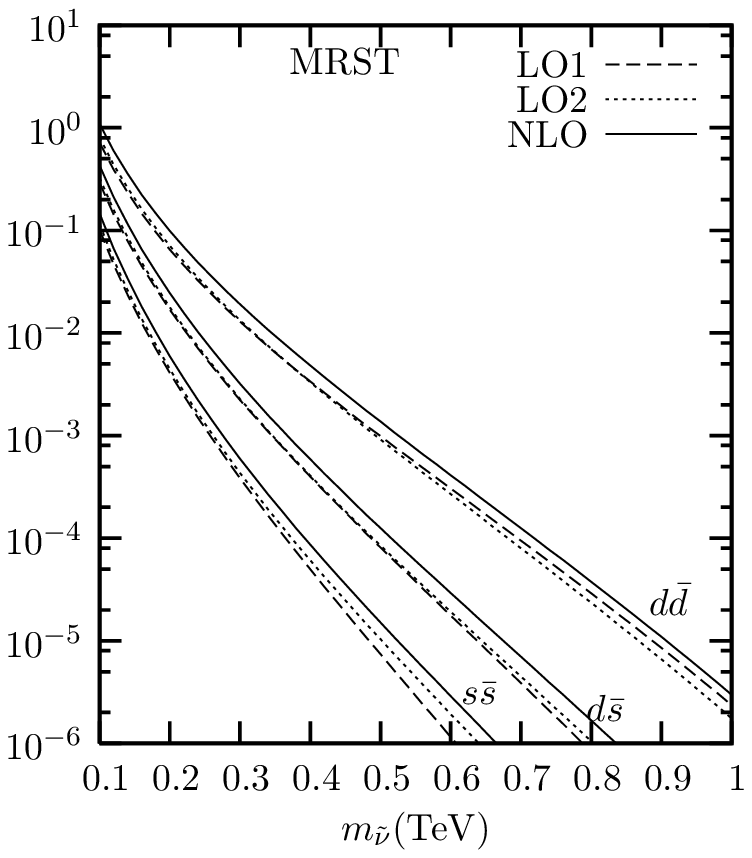}
  \caption{\label{fig:cs_sn_tvt}The cross sections for sneutrino production from various
    initial states at the Tevatron. The cross sections from $s\bar{d}$ initial states is
    obviously equal to those from $d\bar{s}$ initial states for $p\bar{p}$ collision if
    the corresponding $\lambda'$s are equal.}
\end{figure}

\begin{figure}[ht!]
  \centering
  \includegraphics{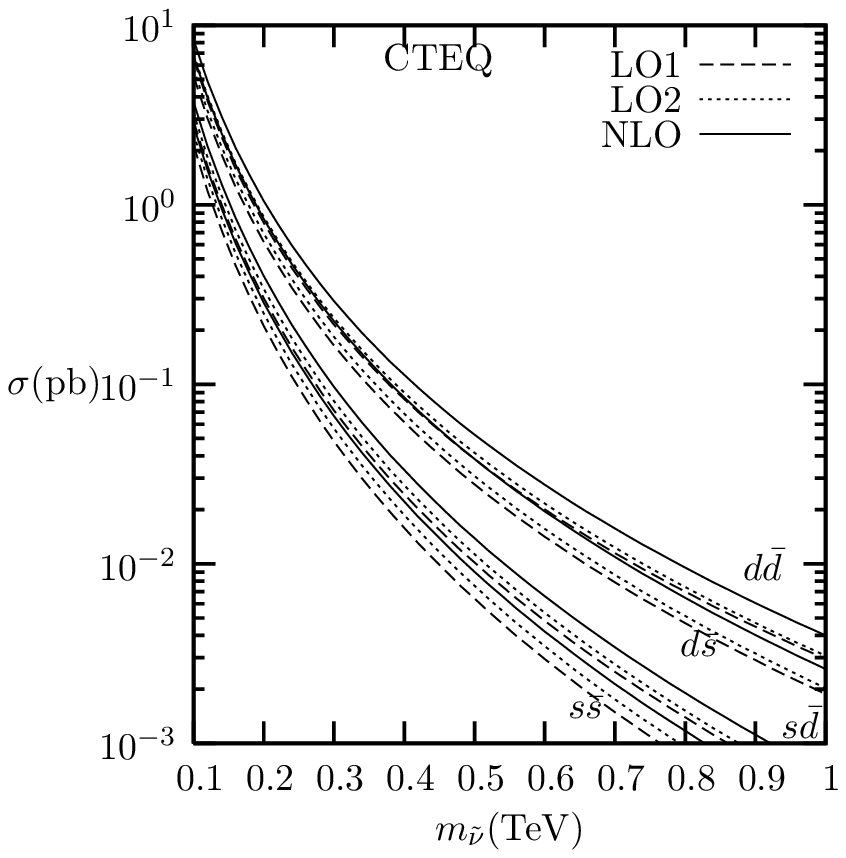}
  \includegraphics{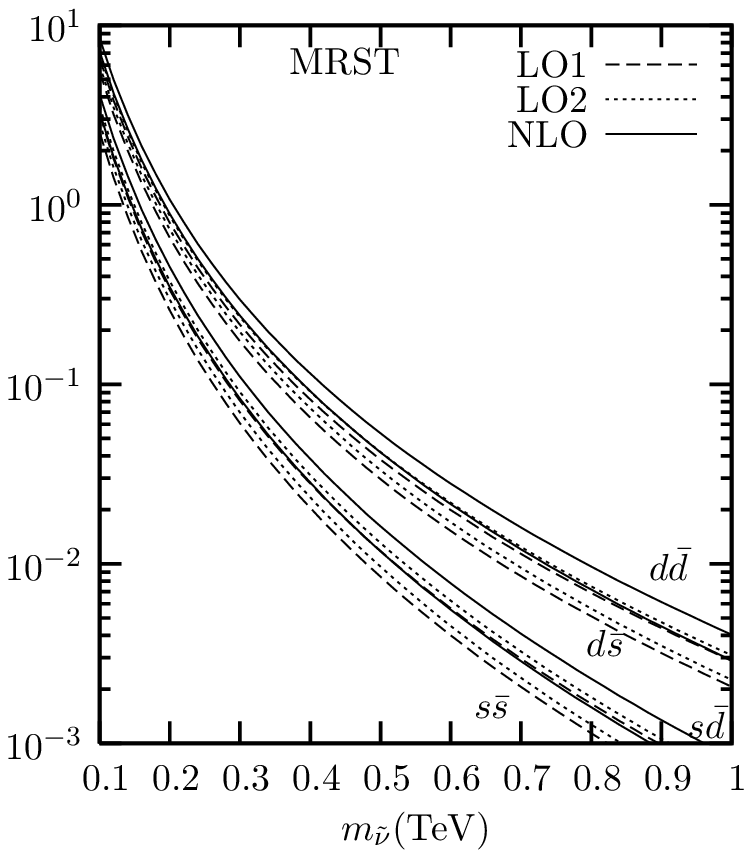}
  \caption{\label{fig:cs_sn_lhc}The cross sections for sneutrino production at the LHC.
    For $pp$ collision $d\bar{s}$ initiated process has larger cross section than
    $s\bar{d}$ initiated one if the couplings are the same.}
\end{figure}

Fig.~\ref{fig:cs_sn_tvt} shows the cross sections for sneutrino production at the
Tevatron as functions of the sneutrino mass. The left and right graphs correspond to
CTEQ6.1 and MRST2004 PDFs, respectively. The solid, dashed and dotted curves represents
to $\sigma_{\text{NLO}}$, $\sigma_{\text{LO1}}$ and $\sigma_{\text{LO2}}$, respectively.
For $p\bar{p}$ collision, since we take the $\lambda'$s to be equal, the cross sections
from $s\bar{d}$ initial states are obviously equal to those from $d\bar{s}$ initial
states. We find that the NLO QCD corrections generally enhance the total cross sections.
The cross sections decrease monotonically with the increasing of the sneutrino mass. From
Fig.~\ref{fig:cs_sn_tvt}, one can see that, for a sneutrino of mass $m=200$~GeV, if we
take $\lambda'=0.02$, the leading order (LO1) and the next-leading-order cross sections
from $d\bar{d}$ channel are 274~fb and 400~fb, respectively. The cross sections from
$d\bar{s}$ channel are 58~fb and 83~fb, while the ones from $s\bar{s}$ channel are 11~fb
and 16~fb, respectively.
The cross sections for the LHC case are shown in Fig.~\ref{fig:cs_sn_lhc}. For $pp$
collider, the cross sections from $d\bar{s}$ and $s\bar{d}$ initial states are no longer
equal even with equal couplings. For a 200~GeV sneutrino and $\lambda'=0.02$, the cross
sections from $d\bar{d}$ channel can reach 3.17~pb (LO1) and 4.2~pb (NLO), respectively.
The cross sections from other channels are relatively smaller, but still remarkable as
the cases at the Tevatron. Our results are consistent with those from
Ref.~\cite{Nucl.Phys.B660.343}.

\begin{figure}[ht!]
  \centering
  \includegraphics{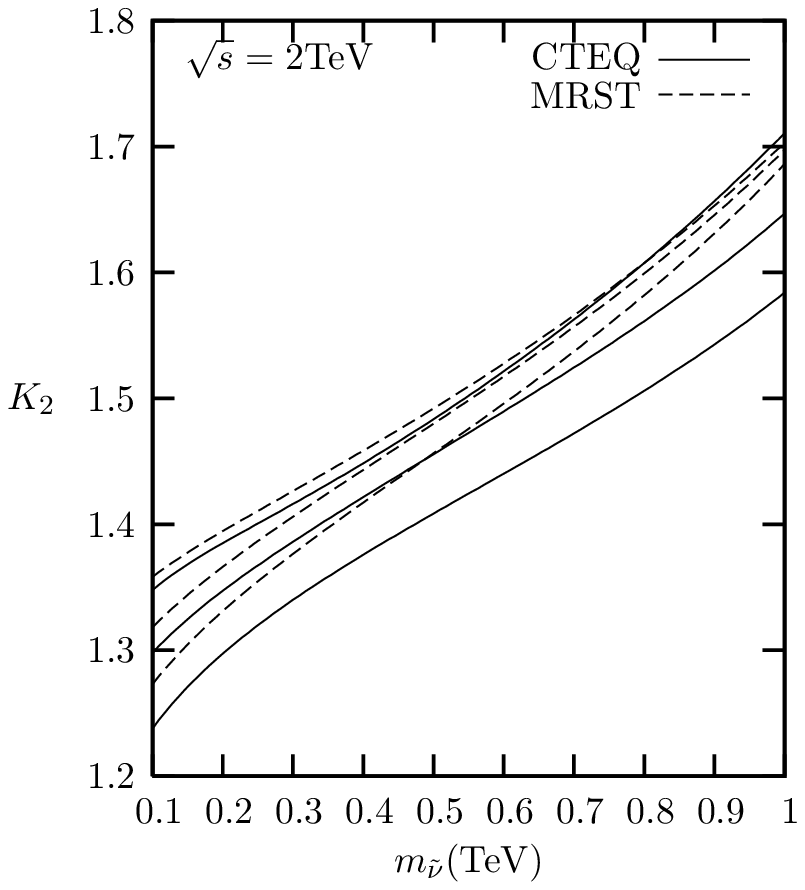}
  \includegraphics{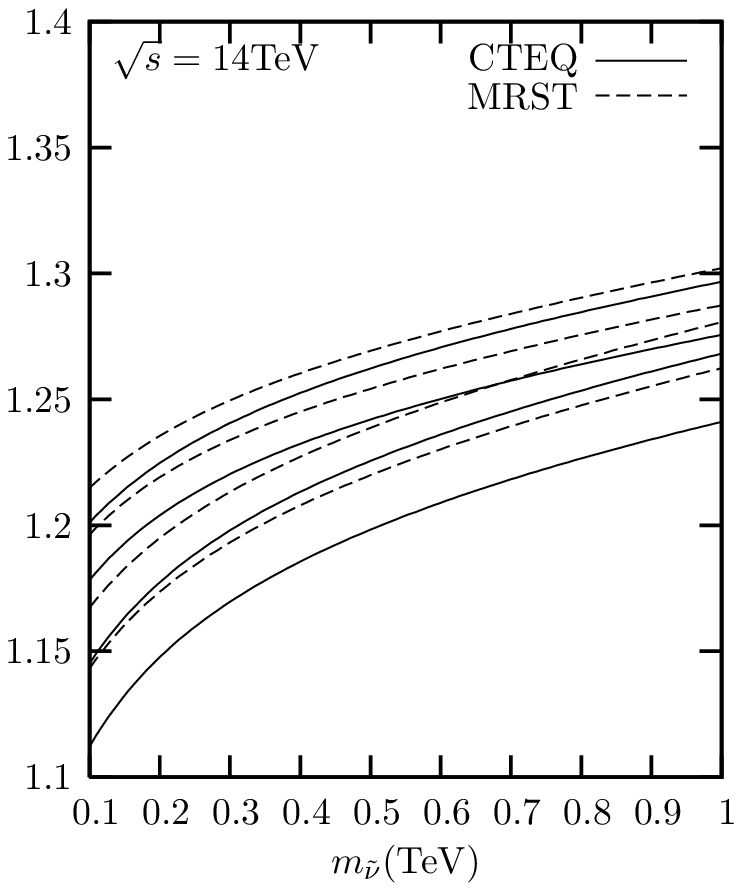}
  \caption{\label{fig:k2_sn}$K_2$ for sneutrino production at the Tevatron (left graph)
    and at the LHC (right graph). The solid and dashed curves correspond to CTEQ and MRST
    PDFs, respectively. For each PDF set, the curves from top to bottom represents
    $d\bar{d}$, $d\bar{s}$, $s\bar{d}$ and $s\bar{s}$ initiated processes, respectively,
    while for the Tevatron case $s\bar{d}$ is omitted.}
\end{figure}

To quantify the enhancement of the total cross sections by the NLO QCD corrections, we
show the $K$ factors in the following. First, we plot $K_2$ as functions of the sneutrino
mass in Fig.~\ref{fig:k2_sn}. The curves are quite similar in spite of different PDF sets
and different initial states, which is just a reflection of the flavour blindness of QCD.
The solid and dashed curves correspond to CTEQ and MRST PDFs, respectively. For each PDF
set, the curves from top to bottom represents $d\bar{d}$, $d\bar{s}$, $s\bar{d}$ and
$s\bar{s}$ initiated processes, respectively, while for the Tevatron case $s\bar{d}$ is
omitted. The ordering of the magnitudes of the $K$ factors of different initial states is
expected, which is due to the different weights of the contributions from gluon initiated
subprocesses.  The corrections increase monotonically with the increasing of sneutrino
mass, and can reach 70 percent at Tevatron and 30 percent at LHC for a 1~TeV sneutrino,
respectively.

\begin{figure}[ht!]
  \centering
  \includegraphics{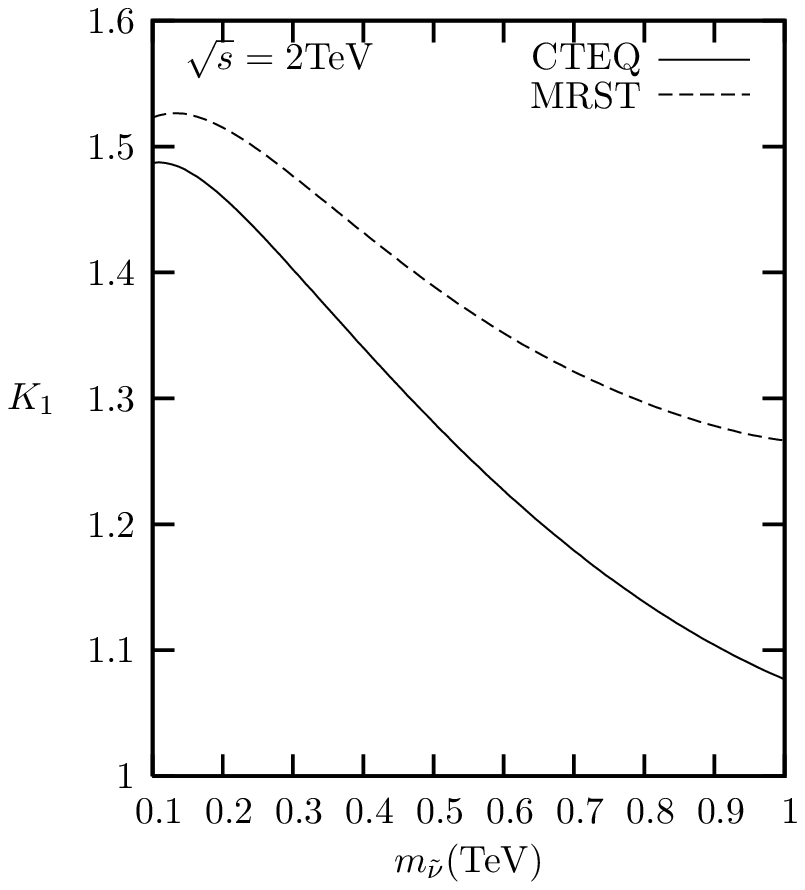}
  \includegraphics{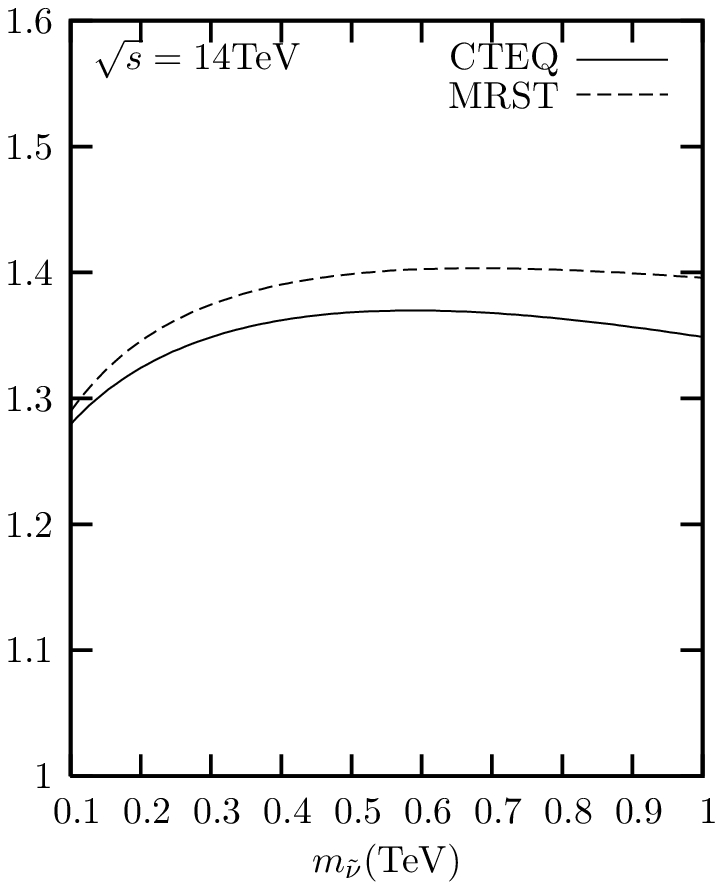}
  \caption{\label{fig:k1_sn_dd}$K_1$ for sneutrino production from $d\bar{d}$ initial
    states at the Tevatron (left graph) and at the LHC (right graph).}
\end{figure}

\begin{figure}[ht!]
  \centering
  \includegraphics{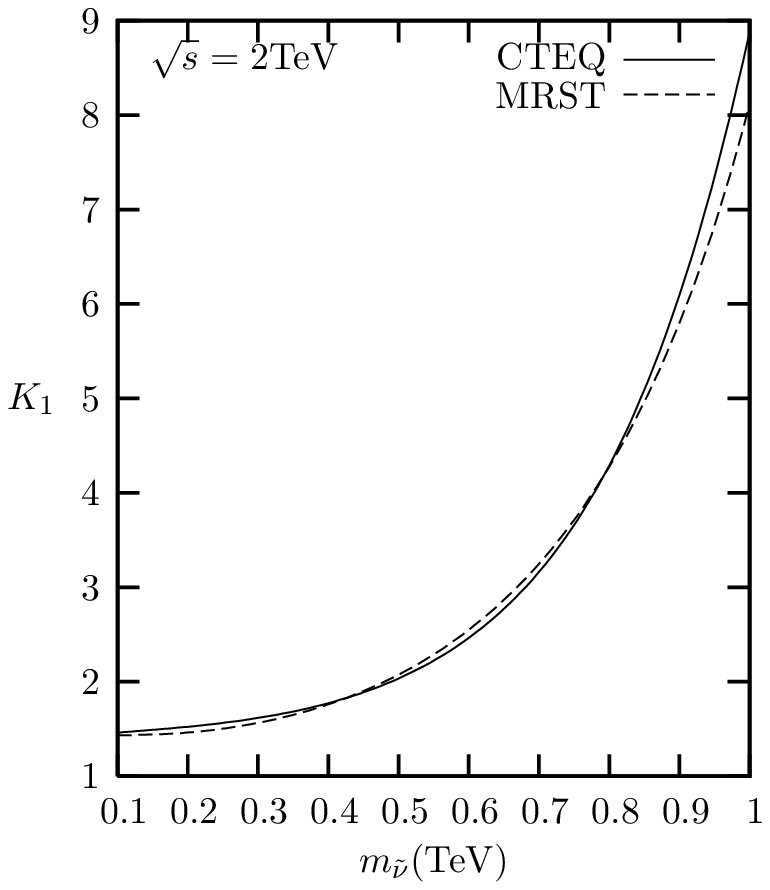}
  \includegraphics{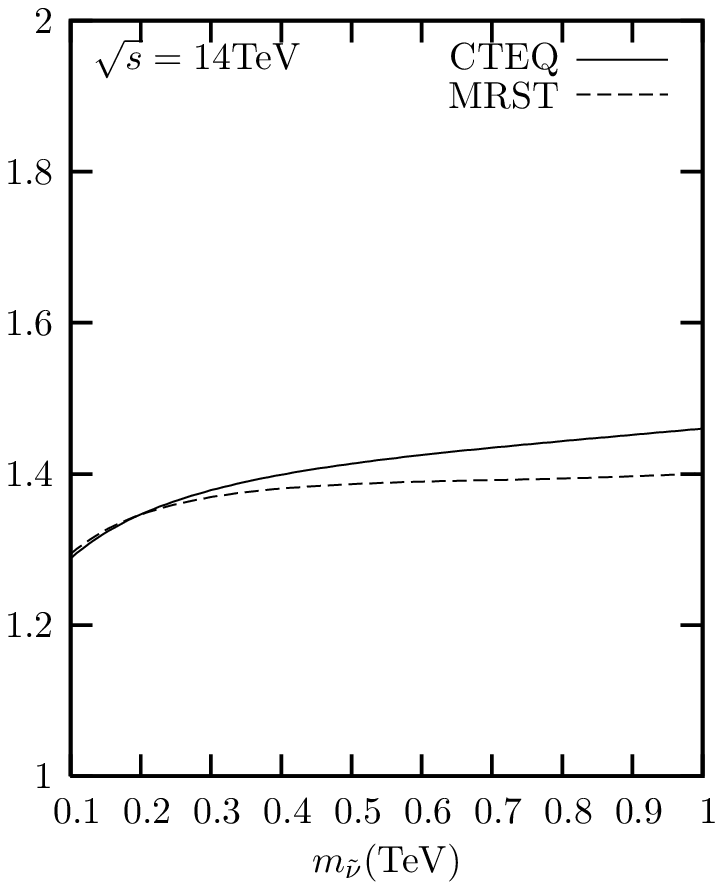}
  \caption{\label{fig:k1_sn_ss}$K_1$ for sneutrino production from $s\bar{s}$ initial
    states at the Tevatron (left graph) and at the LHC (right graph).}
\end{figure}

The case of $K_1$ is a bit more complicated due to the different order of PDFs involved.
In Fig.~\ref{fig:k1_sn_dd} we plot $K_1$ for $d\bar{d}$ initial states as functions of
the sneutrino mass. We find that there is still certain discrepancy between the results
of CTEQ and MRST parton distributions, especially at large $m_{\tilde{\nu}}$ region. This
is mainly due to the difference of the leading order results at region of large momentum
fraction, while the next-to-leading order results of the two PDF sets are quite close to
each other, as shown in Fig.~\ref{fig:cs_sn_tvt}. This confirm the necessity of
calculating the NLO corrections.
For the subprocesses involving strange quark, we only plot $K_1$ for $s\bar{s}$ initial
states. From Fig.~5 in Ref.~\cite{Nucl.Phys.B660.343}, one can see that there was large
discrepancy between the results of CTEQ5 and MRST98 PDF sets. With the updated PDFs, we
find that the discrepancy has been significantly reduced. As shown in
Fig.~\ref{fig:k1_sn_ss}, the two curves agree each other quite well, and are more close
to the CTEQ5 results given in Ref.~\cite{Nucl.Phys.B660.343}.

\begin{figure}[ht!]
  \centering
  \includegraphics{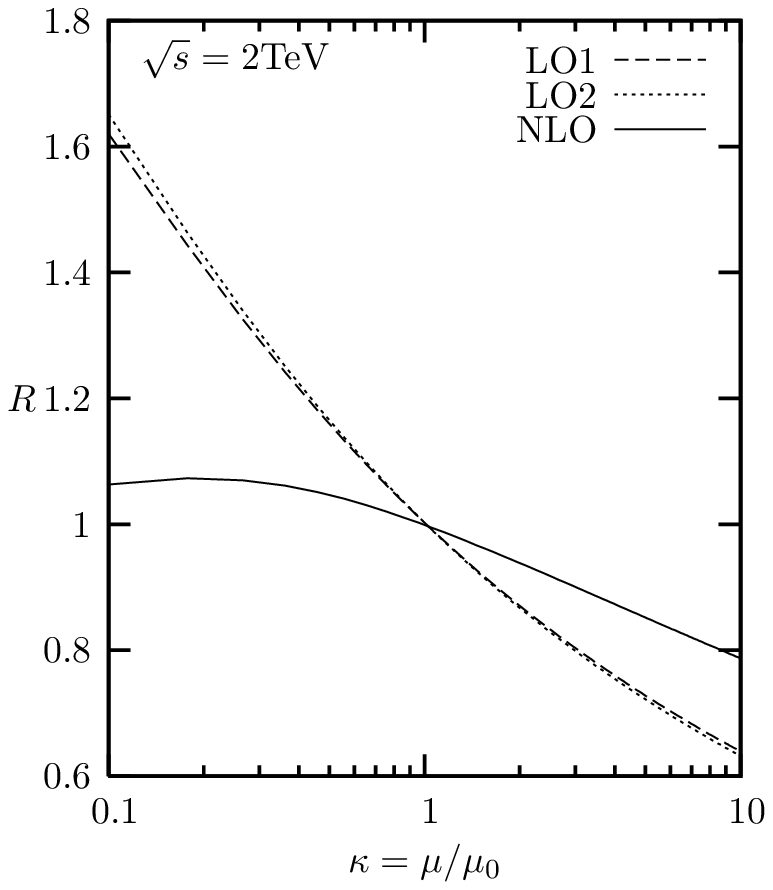}
  \includegraphics{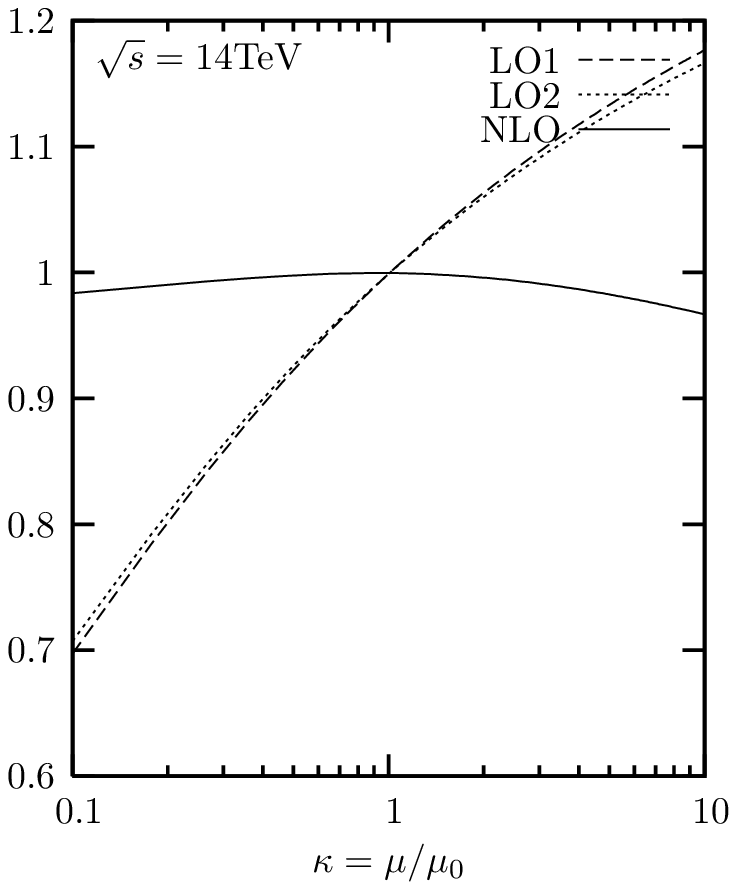}
  \caption{\label{fig:sc_sn}The renormalization and factorization scale dependence of the
    leading order and next-to-leading order cross sections at Tevatron (left graph) and
    at LHC (right graph).}
\end{figure}

We now investigate the dependence of the cross sections on the renormalization and
factorization scales. We will choose the renormalization $\mu_r$ and the factorization
scale $\mu_f$ to be equal, $\mu_r=\mu_f=\kappa\mu_0$, where $\mu_0=m_{\tilde{\nu}}$ is
the mass of the sneutrino. In Fig.~\ref{fig:sc_sn} we plot the ratios of the cross
sections $R=\sigma(\kappa)/\sigma(1)$ as functions of $\kappa$ for
$m_{\tilde{\nu}}=200$~GeV. We only use CTEQ PDFs, since the parametrization of the parton
distribution is irrelevant for the main conclusion here. We show here the results of
$d\bar{d}$ initial states, and the results of other initial states are similar. From the
figure one can see that the dependence on the scales is significantly reduced from LO to
NLO, as we expected.

\subsection{Charged slepton production}

\begin{figure}[ht!]
  \centering
  \includegraphics{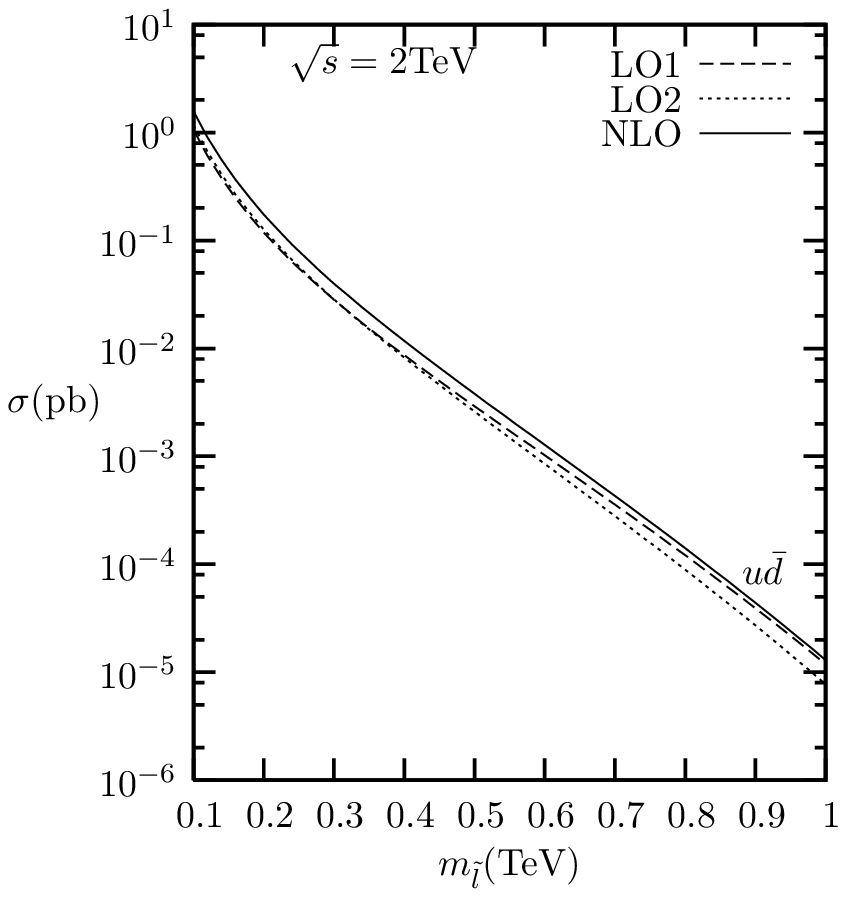}
  \includegraphics{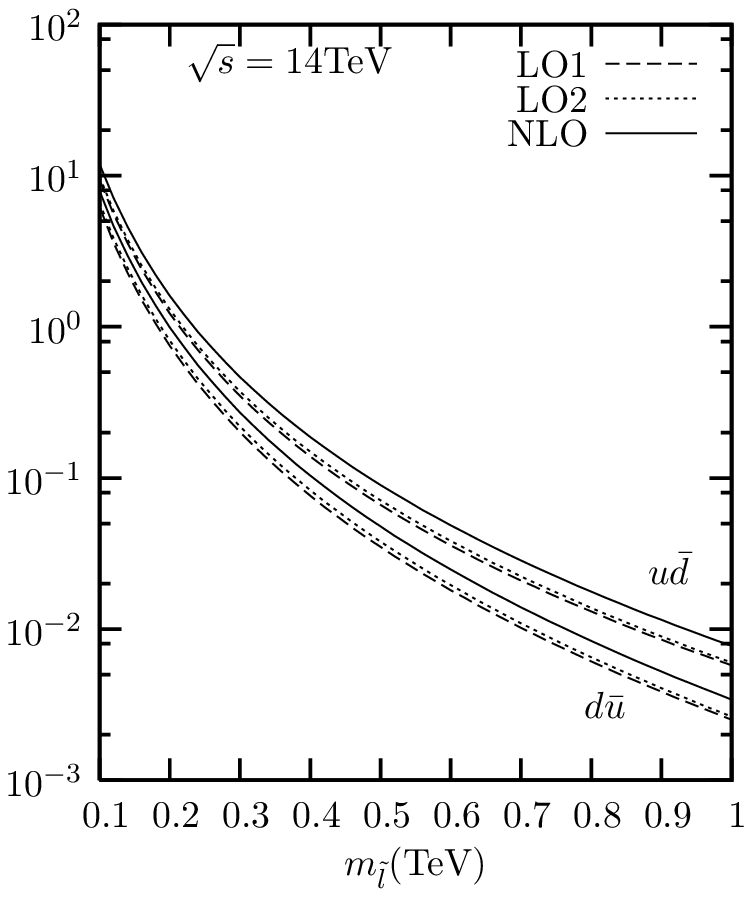}
  \caption{\label{fig:cs_sl}The cross sections for charged slepton production at the
    Tevatron (left) and the LHC (right).}
\end{figure}

\begin{figure}[ht!]
  \centering
  \includegraphics{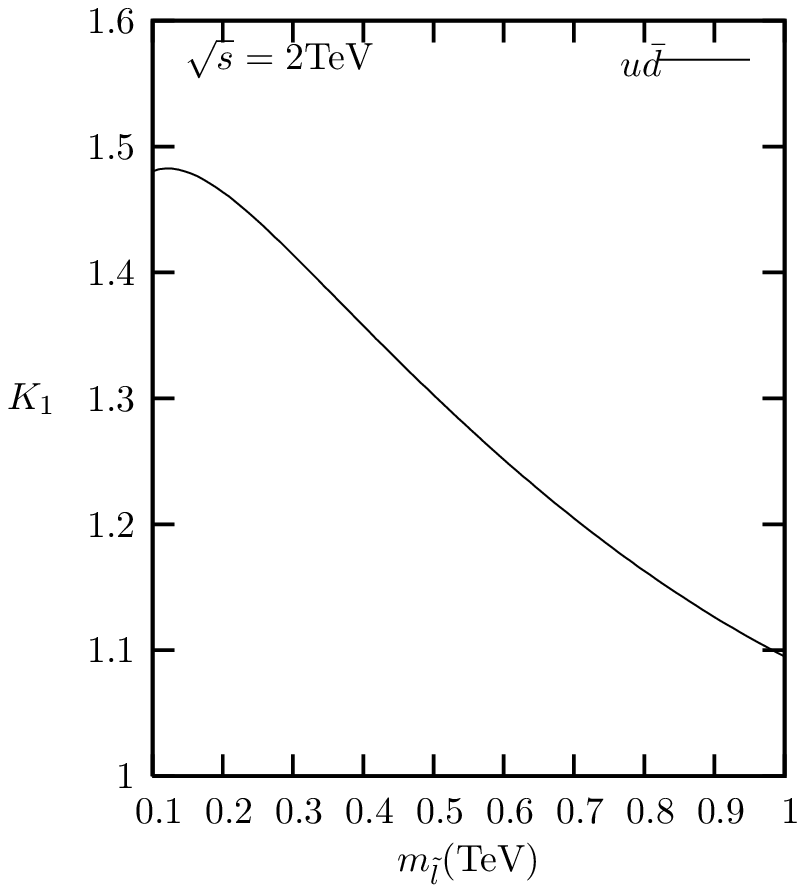}
  \includegraphics{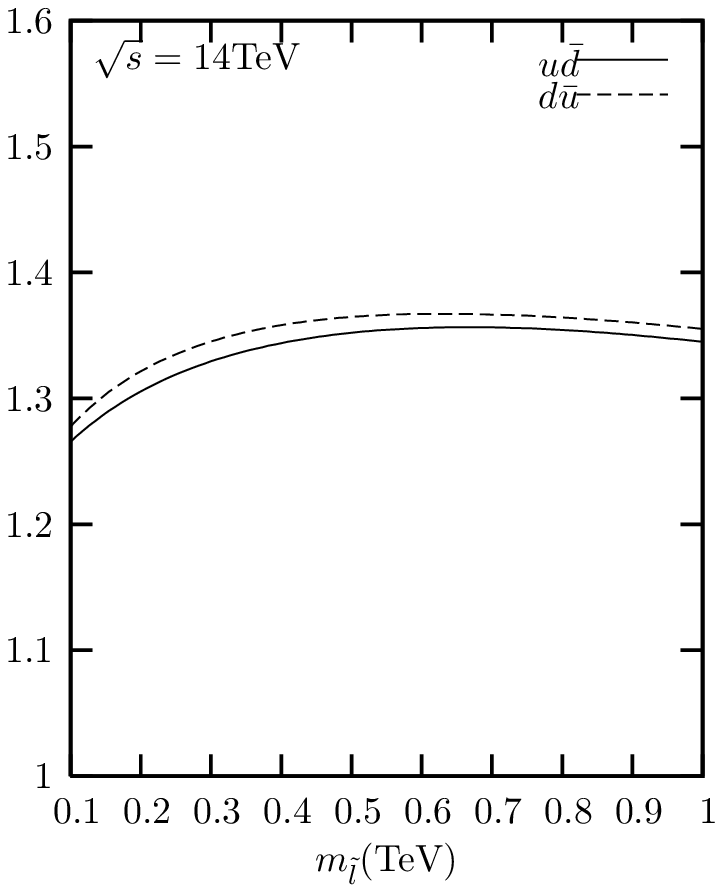}
  \caption{\label{fig:k1_sl}The $K$ factor ($K_1$) for charged slepton production at the
    Tevatron (left) and the LHC (right).}
\end{figure}

We now turn to the production of charged sleptons. The possible initial states now become
$u\bar{d}$ and $u\bar{s}$ for sleptons with positive charge, $d\bar{u}$ and $s\bar{u}$
for ones with negative charge, respectively. For simplicity, we will only consider
contributions from $u\bar{d}$ and $d\bar{u}$ initial states, and only use the CTEQ6.1
PDF. The situation is much similar to the production of sneutrinos, so we do not discuss
any details here and just show the cross sections in Fig.~\ref{fig:cs_sl} and the $K$
factors in Fig.~\ref{fig:k1_sl}, respectively.



\section{Transverse momentum distribution}\label{sec:qt}

In this section we investigate the transverse momentum distribution of the slepton. At
Born level the slepton is kept at zero $q_T$ due to momentum conservation and the
distribution is proportional to $\delta^2(\vec{q}_T)$. Thus the leading order
distribution at non-zero $q_T$ belongs to $\mathcal{O}(\alpha_s)$, where momentum
conservation is retained by the additional parton emitted. The distribution can be
obtained by integrate out $y$ from the differential cross section given in
Section~\ref{sec:nlo}. However, this fixed order result is only valid when $q_T$ is not
too small compared with the mass of the slepton $m$. If $q_T \ll m$, the corresponding
parton emitted would be either soft or collinear to one of the initial partons. In
consequence, large logarithms like $\ln(m^2/q_T^2)$ will appear and will dominate the
cross section for sufficiently small $q_T$. In general, there should be double logarithms
for each gluon attached to the initial quarks due to the overlap of soft region and
collinear region. As a result, the perturbative expansion would be controlled by
$\alpha_s\ln^2(m^2/q_T^2)$ rather than $\alpha_s$. The convergence of the perturbation
series will be spoiled if $\alpha_s\ln^2(m^2/q_T^2)$ approaches unity.

The problem arising at small $q_T$ can also be seen from the fact that the fixed order
cross section is singular as $q_T \to 0$. For later use, we define the asymptotic part of
the differential cross section at small $q_T$ to be the terms which are at least as
singular as $1/q_T^2$ when $q_T \to 0$. At $\mathcal{O}(\alpha_s)$, the asymptotic
expression can be isolated from the expressions obtained in section~\ref{sec:nlo}:
\begin{align}
  \label{eq:asym}
  \frac{\diff\sigma}{\diff{q_T^2}\diff{y}}(\text{asym}) &= \frac{1}{2} \sigma_0
  \frac{\alpha_s}{\pi} \frac{1}{q_T^2} \biggl\{ f_{\alpha/A}(x_1^0,\mu_f)
    f_{\beta/B}(x_2^0,\mu_f) \left( 2C_F \ln\frac{m^2}{q_T^2} - 3C_F \right) \biggr.
  \notag \\
  &\qquad + \biggl. (P\circ{f})_{\alpha/A}(x_1^0,\mu_f) f_{\beta/B}(x_2^0,\mu_f) +
      f_{\alpha/A}(x_1^0,\mu_f) (P\circ{f})_{\beta/B}(x_2^0,\mu_f) \biggr\},
\end{align}
where $\sigma_0=\displaystyle{\frac{\pi}{12s}}\lambsq$.

\subsection{Resummation of large logarithms}

In order to make use of the perturbation theory with the existence of large logarithms at
each order, one must reorganize the perturbative expansion to resum the large terms. It
was first shown by Dokshitzer, Diakonov and Troyan (DDT) \cite{Phys.Rep.58.269} that in
the double leading logarithm approximation (DLLA), terms like
$\alpha_s^n\ln^{2n-1}(m^2/q_T^2)/q_T^2$ can be resummed into a Sudakov form factor.
However, this result relies on the assumption that the emitted gluons are both soft and
collinear and their transverse momenta are strongly ordered.  With the over-constrained
phase space, the resulting distribution is over-suppressed at small $q_T$. This implies
that the subleading logarithms are also important and need to be resummed too. The
subleading logarithms correspond to the phase space configurations in which some of the
emitted gluons are not so soft or collinear. In this case, the transverse momentum
conservation must be imposed, which is implemented in the $b$-space formalism introduced
by Parisi and Petronzio \cite{Nucl.Phys.B154.427}. Collins and Soper \cite{colsop}
improved the $b$-space formalism to resum all the terms like $\alpha_s^nL^r$
($r=0,...,2n-1$) based on the renormalization group equation (RGE) method, where $L$
represents the large logarithms. In this framework, Collins, Soper and Sterman (CSS)
\cite{Nucl.Phys.B250.199} derived a resummation formula for the transverse momentum
distributions of the vector bosons produced in Drell-Yan processes. This formalism, often
refered to as the CSS formalism, has also been applied to many other processes.

In the CSS formalism, the differential cross section we are considering can be written as
\begin{equation}
  \label{eq:total}
  \frac{\diff\sigma}{\diff{q_T^2}\diff{y}} = \frac{\diff\sigma}{\diff{q_T^2}\diff{y}}
  (\text{resum}) + Y(q_T,m,x_1^0,x_2^0),
\end{equation}
where the resummed part can be expressed as an inverse Fourier transformation
\begin{align}
  \label{eq:resum}
  \frac{\diff\sigma}{\diff{q_T^2}\diff{y}} (\text{resum}) &= \sum_{\alpha,\beta}
  \frac{1}{2} \sigma_0 \frac{1}{2\pi} \int \diff^2\vec{b} \exp \left( i\vec{b} \cdot
    \vec{q}_T \right) W_{\alpha\beta}(b,m,x_1^0,x_2^0)
  \notag \\
  &= \sum_{\alpha,\beta} \frac{1}{2} \sigma_0 \int_0^\infty b \diff{b} J_0(b q_T)
  W_{\alpha\beta}(b,m,x_1^0,x_2^0),
\end{align}
where $\vec{b}$ is the impact parameter conjugating to $\vec{q}_T$, $J_0$ is zero order
Bessel function of the first kind, and
\begin{align}
  W_{\alpha\beta}(b,m,x_1^0,x_2^0) &= \tilde{f}_{\alpha/A}(x_1^0,C_3/b)
  \tilde{f}_{\beta/B}(x_2^0,C_3/b) \notag
  \\
  &\qquad \times \exp \left\{ - \int_{C_1^2/b^2}^{C_2^2m^2}
    \frac{\diff\bar{\mu}^2}{\bar{\mu}^2} \left[ \ln\frac{C_2^2m^2}{\bar{\mu}^2}
      A(\alpha_s(\bar{\mu})) + B(\alpha_s(\bar{\mu})) \right] \right\}.
\end{align}
Here $C_i (i=1,2,3)$ are constants of order 1 which are by convention
\cite{Nucl.Phys.B250.199} chosen to be
\begin{equation}
  C_1 = C_3 = 2 e^{-\gamma_E} \equiv b_0, \, C_2 = 1,
\end{equation}
and $\tilde{f}$ is the convolution of the PDFs and the coefficient functions $C$
\begin{equation}
  \tilde{f}_{\alpha/h}(x,\mu) = \sum_{\gamma} \int_x^1 \frac{\diff{z}}{z}
  C_{\alpha\gamma}(z,\alpha_s(\mu)) f_{\gamma/h}(x,\mu),
\end{equation}
and the coefficients $A$, $B$ and $C$ can be expanded to series in $\alpha_s$
\begin{align}
  A(\alpha_s) &= \sum_{n=1}^\infty A^{(n)} \left( \frac{\alpha_s}{\pi} \right)^n,
  \\
  B(\alpha_s) &= \sum_{n=1}^\infty B^{(n)} \left( \frac{\alpha_s}{\pi} \right)^n,
  \\
  C_{\alpha\beta}(z,\alpha_s) &= \sum_{n=0}^\infty C_{\alpha\beta}^{(n)}(z) \left(
    \frac{\alpha_s}{\pi} \right)^n,
\end{align}
and they can be calculated order by order in perturbative theory. The lowest order
coefficients can be extracted from the asymptotic expression above
\begin{align}
  A^{(1)} &= C_F = \frac{4}{3} , \qquad B^{(1)} = - \frac{3}{2} C_F = -2 ,
  \\
  C_{\alpha\beta}^{(0)}(z) &= \delta_{\alpha\beta} \delta(1-z).
\end{align}
With these coefficients, we can actually sum up all terms like $\alpha_s^nL^{2n-1}$ and
$\alpha_s^nL^{2n-2}$.

The another term in Eq.~(\ref{eq:total}), the $Y$ term, is the remaining contributions
which are not resummed. Since it contains no large logarithms, it can be reliably
calculated in perturbation theory. In fact, according to the definition of the asymptotic
part mentioned above, one can immediately see that
\begin{equation}
  Y = \frac{\diff\sigma}{\diff{q_T^2}\diff{y}} (\text{pert}) -
  \frac{\diff\sigma}{\diff{q_T^2}\diff{y}} (\text{asym}).
\end{equation}
However, the resummed part is still not able to be calculated perturbatively. The reason
is that in Eq.~(\ref{eq:resum}), the integral over the impact parameter $b$ extends to
infinity, while the integrand involves the strong coupling constant $\alpha_s$ and the
PDFs at scale $b_0/b$, where they are not well defined if $b$ is large enough so that
$b_0/b$ enters non-perturbative region.

Many prescriptions have been developed regarding this problem, and we will show two
approaches here. Collins, Soper and Sterman, in their original paper
\cite{Nucl.Phys.B250.199}, suggested that one can use a cut-off $b_{\text{max}}$ and
regard the effects from $b>b_{\text{max}}$ as non-perturbative input. Practically, they
replacing $W(b)$ in Eq.~(\ref{eq:resum}) by
\begin{equation}
  \widetilde{W}(b) = W(b_*) F_{\text{NP}}(b),
\end{equation}
where
\begin{equation}
  b_* = \frac{b}{\sqrt{1+(b/b_{\text{max}})^2}},
\end{equation}
and $F_{\text{NP}}(b)$ parameterizes the non-perturbative effects. Since $b_*$ never
exceeds $b_{\text{max}}$, $W(b_*)$ can be calculated perturbatively, and the theoretical
uncertainty mainly relies on the function $F_{\text{NP}}$. Recently, Landry, Brock,
Nadolsky and Yuan (BLNY) \cite{Phys.Rev.D67.073016} proposed the form
\begin{equation}
  F_{\text{NP}} = \exp \left\{ - b^2 \left[ g_1 + g_2 \ln\frac{m}{2Q_0} +
      g_1g_3\ln(100x_1^0x_2^0) \right] \right\}.
\end{equation}
They take $b_{\text{max}} = 0.5 \text{GeV}^{-1}$, $Q_0 = 1.6 \text{GeV}$ and the
parameters $g_i (i=1,2,3)$ are fitted to the available Drell-Yan data, which are given by
\begin{equation}
  g_1 = 0.21, \qquad g_2 = 0.68, \qquad g_3 = -0.60.
\end{equation}

One of the largest disadvantages of the $b_*$ prescription is that it alters the $W$
function in the perturbative region $b<b_{\text{max}}$. Regarding this, Qiu and Zhang
(QZ) \cite{Phys.Rev.D63.114011} proposed another prescription so that $W(b)$ is kept
unchanged for $b<b_{\text{max}}$.  Namely, they chose
\begin{equation}
  \widetilde{W}(b) =
  \begin{cases}
    W(b) & b \leq b_\text{max},
    \\
    W(b_\text{max}) F_{\text{NP}}(b) & b > b_\text{max},
  \end{cases}
\end{equation}
where $F_{\text{NP}}$ takes the form
\begin{equation}
  F_{\text{NP}}(b) = \exp \left\{ - \ln\frac{m^2b_{\text{max}}^2}{b_0^2} \left[ g_1
      \left( (b^2)^\alpha - (b_\text{max}^2)^\alpha \right) + g_2 ( b^2 -
      b_{\text{max}}^2 ) \right] - \bar{g}_2 ( b^2 - b_{\text{max}}^2 ) \right\}.
\end{equation}
The parameters $g_1$ and $\alpha$ are fixed by the continuity of the first and second
derivative of $\widetilde{W}$ at $b=b_\text{max}$, while $g_2$ and $\bar{g}_2$ are
determined from experiments. In the numerical evaluation, we will use both above methods
and compare their results.

\subsection{Numerical results}

In the numerical evaluation of the transverse momentum distribution, we take the
renomalization and factorization scale for the fixed order expressions to be
$\mu_r=\mu_f=\sqrt{m^2+q_T^2}/2$. For simplicity, we only show the results for $d\bar{d}$
initial states as an example. The results for other initial states are similar. We use
the CTEQ PDFs throughout this section.

According to the previous analysis, the differential cross section with respect to
$q_T^2$ and $y$ can be formally written as
\begin{equation} \label{eq:total2}
  \frac{\diff\sigma}{\diff{q_T^2}\diff{y}} (\text{total}) =
  \frac{\diff\sigma}{\diff{q_T^2}\diff{y}} (\text{resum}) +
  \frac{\diff\sigma}{\diff{q_T^2}\diff{y}} (\text{pert}) -
  \frac{\diff\sigma}{\diff{q_T^2}\diff{y}} (\text{asym}).
\end{equation}
At small $q_T$, the asymptotic part and the perturbative part cancel each other, and the
resummed part dominates the distribution. At large $q_T$, the difference between the
resummed part and the asymptotic part belongs to higher orders in $\alpha_s$, so the
perturbative predictions are recovered. Since the perturbative part and the asymptotic
part can be reliably calculated within perturbation theory, the only ambiguity comes from
the resummed part due to the non-perturbative issue. So we shall compare the results of
the different approaches to the non-perturbative parametrization for the reliable
predictions.

\begin{figure}[ht!]
\centering
\includegraphics{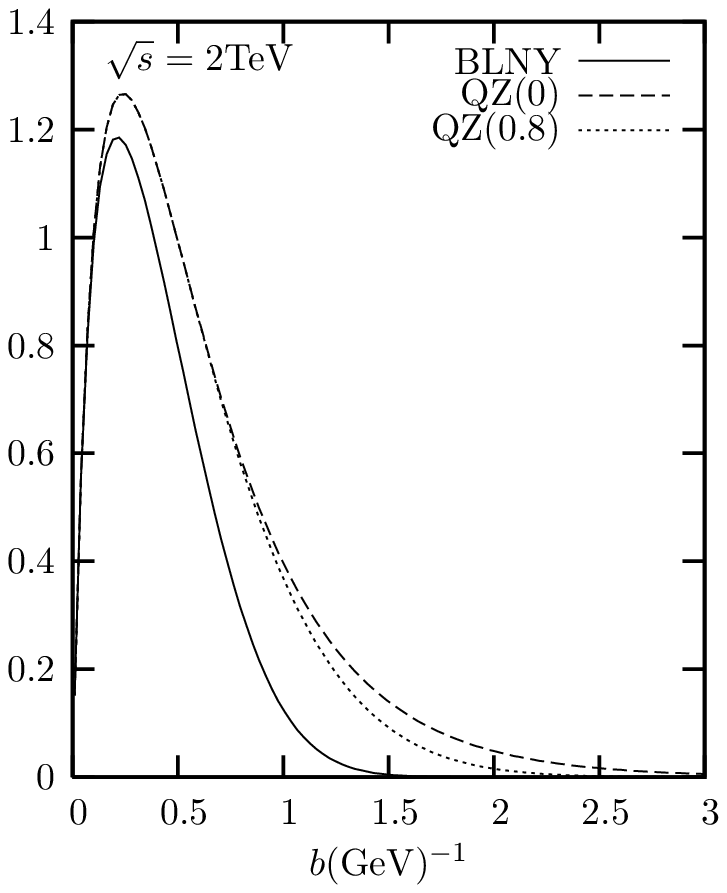}
\includegraphics{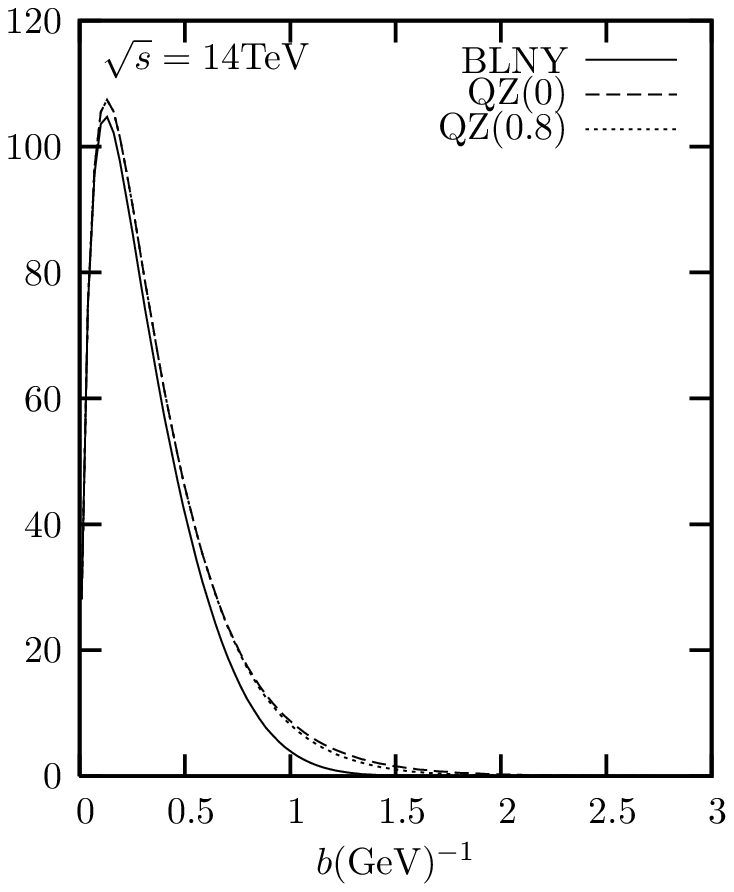}
\caption{\label{fig:bw}Integrand of the $b$-integral (without the Bessel function) as a
  function of $b$ for $m=200$~GeV and $y=0$ at the Tevatron (left) and the LHC (right).}
\end{figure}

With the two $b$-space prescriptions at hand (BLNY and QZ), we first compare the
integrand of the Bessel transformation $b\widetilde{W}(b)$. We plot this function for a
sneutrino with mass $m=200$~GeV and rapidity $y=0$ in Fig.~\ref{fig:bw}. The solid curves
represent the BLNY results, while the dashed and the dotted curves correspond to the the
QZ prescription with two choices of $g_2$: $g_2=0$ (no power corrections) and
$g_2\log(m^2b_{\text{max}}^2/b_0^2)=0.8\text{GeV}^2$ (fitted power corrections),
respectively. The parameter $\bar{g}_2$ is always taken to be zero.
From Fig.~\ref{fig:bw}, we can see that at the Tevatron, the BLNY parametrization notably
changes the shape of the function for $b<b_{\text{max}}$, and shows a sharper decrease
than the one of QZ. As a result, the resummed differential cross sections for the two
approaches will differ at small $q_T$ (at larger $q_T$ the impact of large $b$ region
gets smaller due to the Bessel function). On the other hand, the QZ parametrization
itself is insensitive to the value of $g_2$. At the LHC, as shown in the right graph of
Fig.~\ref{fig:bw}, the difference is quite small, and the resummed cross sections are
expected to agree to each other.  Furthermore, the shape of the function is ``narrower''
at the LHC than the one at the Tevatron, i.e., the contributions from large $b$ region to
the integral at the LHC are smaller than the ones at the Tevatron. So we conclude that
the predictions at the LHC is more reliable than the ones at the Tevatron.
\begin{figure}[ht!]
\centering
\includegraphics{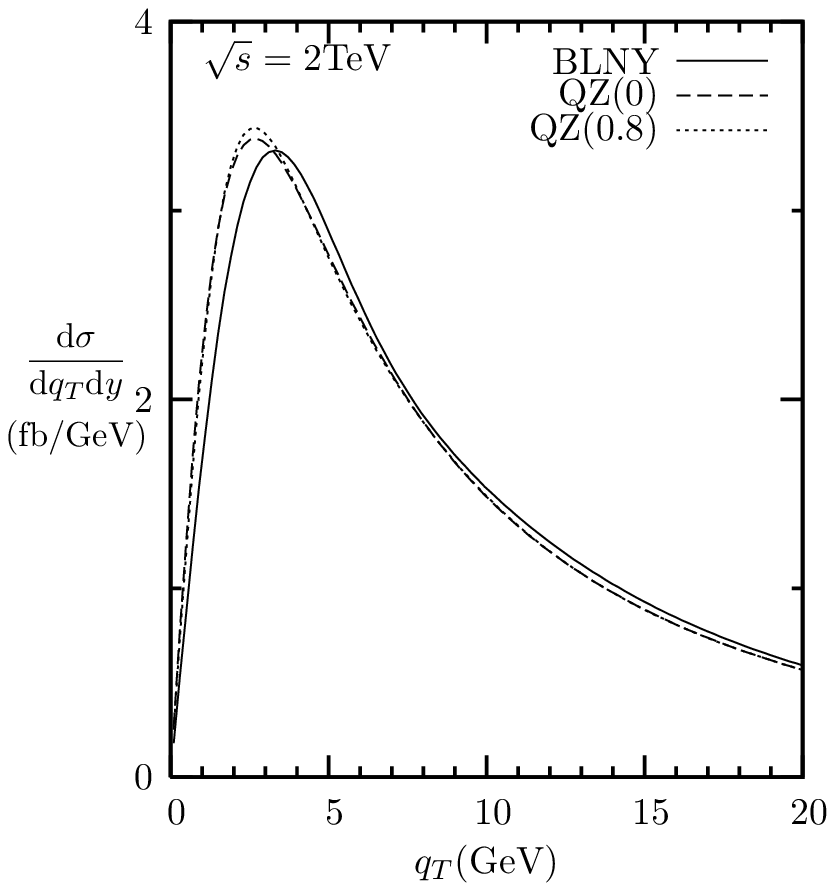}
\includegraphics{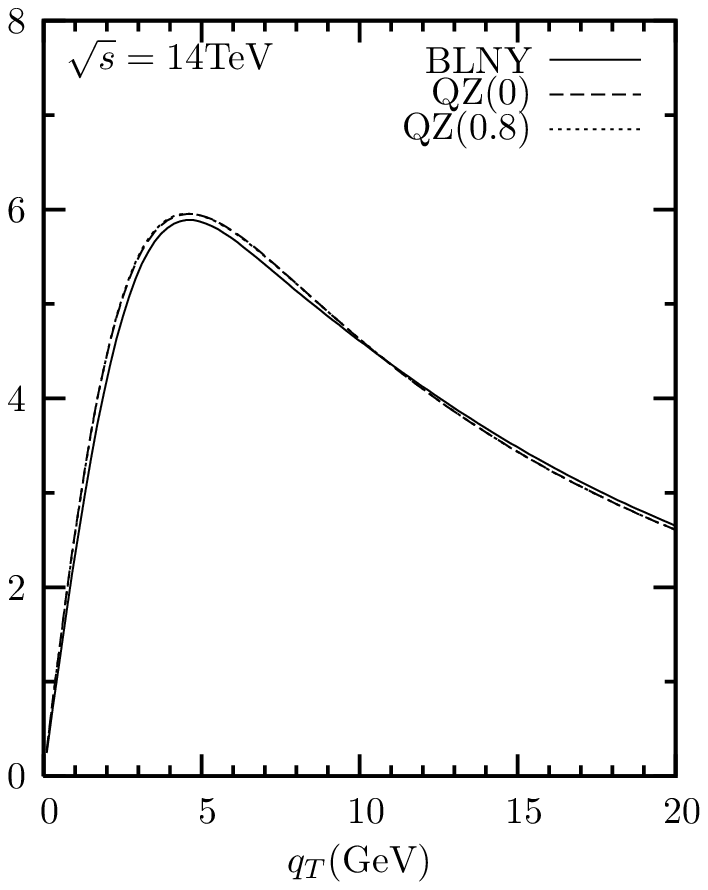}
\caption{\label{fig:qt_cmp}Comparison of the resummed differential cross sections for
  $m=200$~GeV and $y=0$ at the Tevatron (left) and the LHC (right).}
\end{figure}
In Fig.~\ref{fig:qt_cmp}, we plot the corresponding differential cross sections according
to the two approaches. As discussed above, the results are slightly different at Tevatron
and agree quite well at LHC. In the following calculations, we will adopt the QZ
prescription and take $g_2=0$.

\begin{figure}[ht!]
\centering
\includegraphics{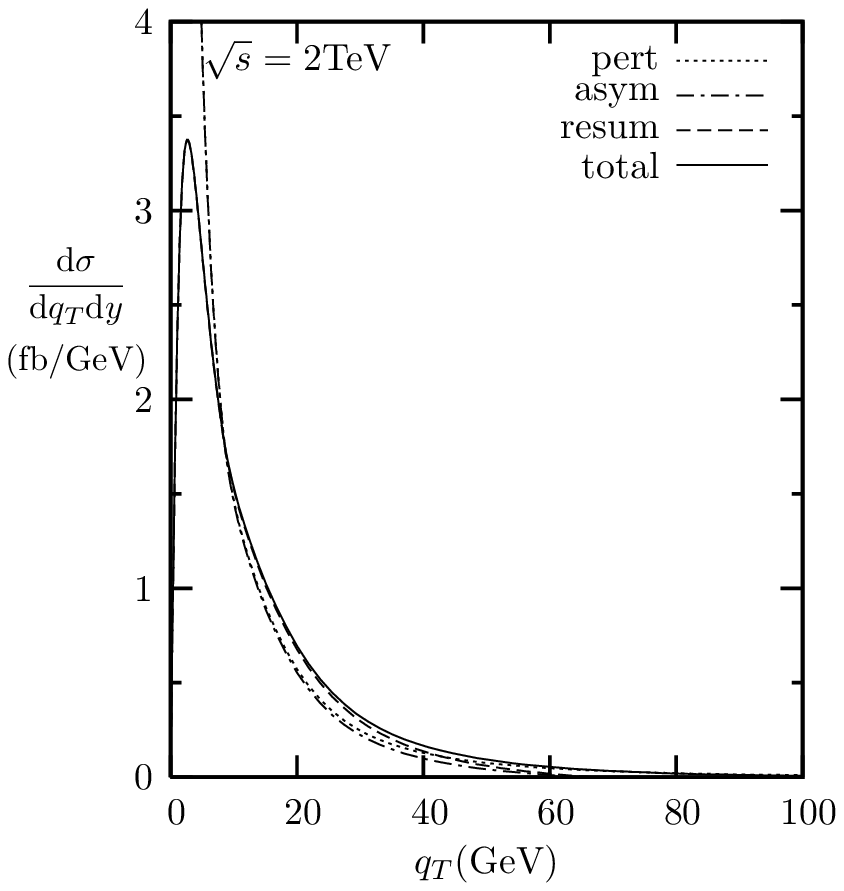}
\includegraphics{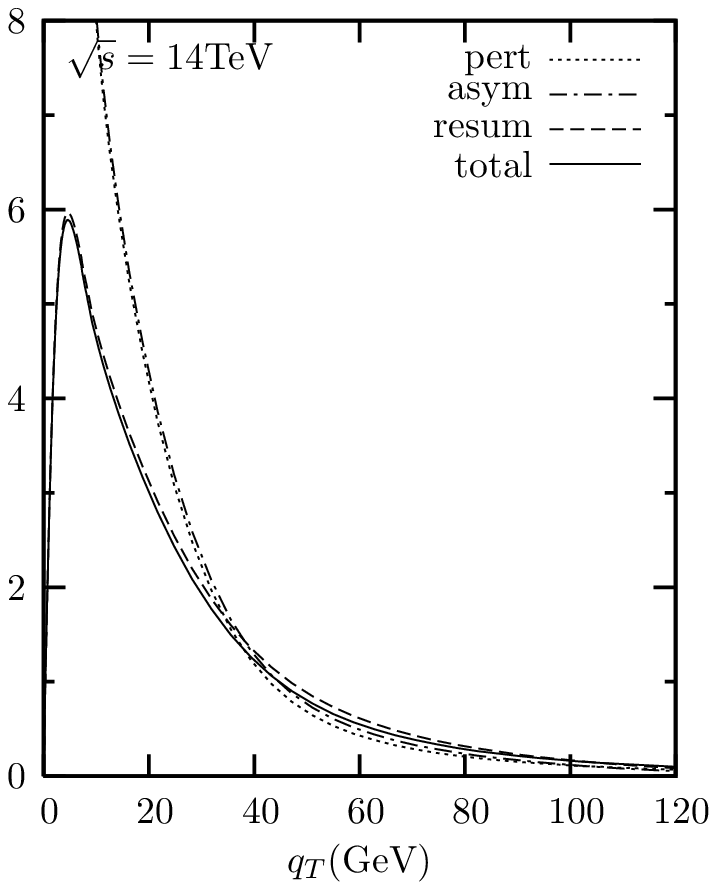}
\caption{\label{fig:qt_y0_200}The distribution with respect to the transverse momentum
  and the rapidity of the sneutrino from $d\bar{d}$ channel for central rapidity $y=0$ at
  the Tevatron (left graph) and the LHC (right graph). The mass of the sneutrino is taken
  to be 200~GeV.}
\end{figure}
In Fig.~\ref{fig:qt_y0_200}, we plot the various parts of the differential cross section
in Eq.~(\ref{eq:total2}) for rapidity $y=0$ at the Tevatron and the LHC. The mass of the
sneutrino is taken to be 200~GeV. The solid, dashed, dotted and dot-dashed curves
correspond to the total result, the resummed part, the perturbative part and the
asymptotic part, respectively. Quantitatively, the perturbative and the asymptotic cross
sections agree very well at small transverse momentum. On the other hand, the resummed
and the asymptotic part are not cancelled completely at high $q_T$ due to the higher
order effects included in the resummed one, so that the total one and the perturbative
one will differ at large $q_T$. This can be considered as the theoretical uncertainties.
In principle, one can return to the perturbative result for $q_T>q_T^{\text{cut}}$, where
$q_T^{\text{cut}}$ is arbitrarily chosen in the intermediate $q_T$ region.  However, in
order to make the transition smooth, one must introduce some kinds of matching procedure
which could also lead to uncertainties. In our work, we will use the resummed cross
section from small $q_T$ to large $q_T$.

\begin{figure}[ht!]
\centering
\includegraphics{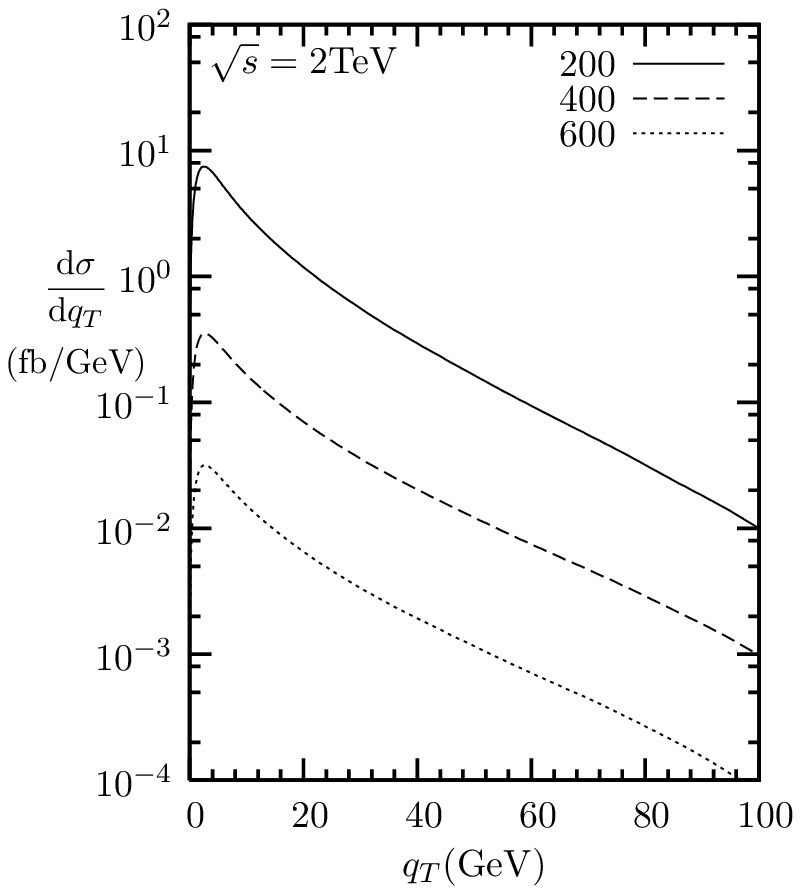}
\includegraphics{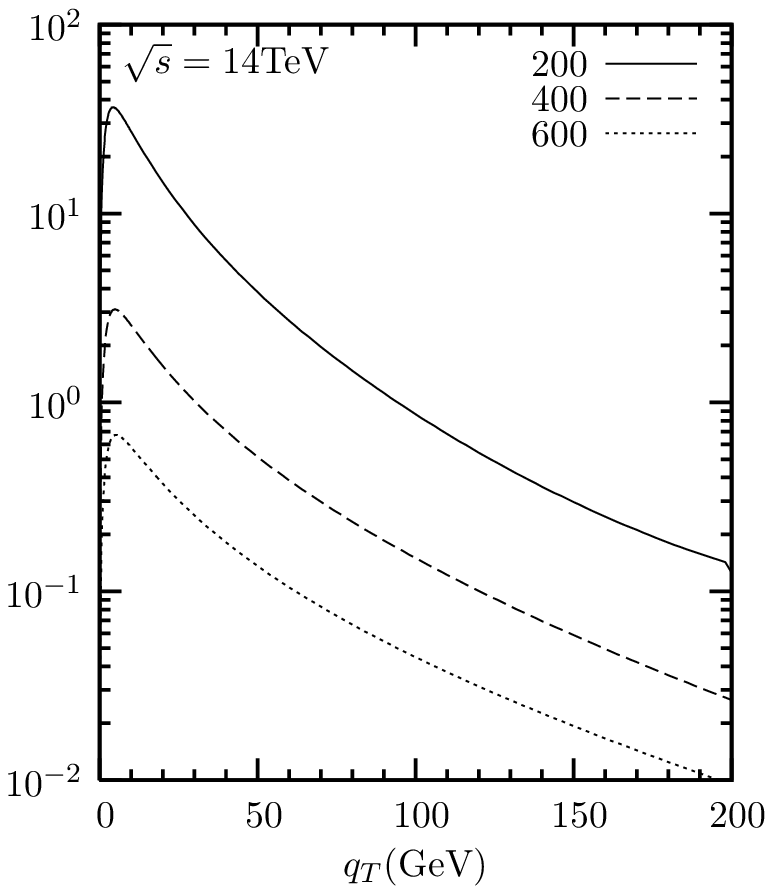}
\caption{\label{fig:qt}The transverse momentum distribution of the sneutrino from
  $d\bar{d}$ channel at the Tevatron (left graph) and the LHC (right graph). The mass of
  the sneutrino is taken to be 200~GeV, 400~GeV, 600~GeV, as labeled in the figure.}
\end{figure}

Having settled down the technical issues, we now turn to show the predictions for the
transverse momentum distributions of the slepton. For the $y$ integral, the kinematic
range should be
\begin{equation}
  |y| \leq \acosh \left( \frac{1+\tau}{2\sqrt{\tau+q_T^2/s}} \right).
\end{equation}
In Fig.~\ref{fig:qt} we plot the transverse momentum distribution of the sneutrino from
the $d\bar{d}$ channel at the Tevatron and the LHC separately. The mass of the sneutrino
is taken to be 200~GeV, 400~GeV, 600~GeV, respectively. The peaks of the distribution
appear at about 3~GeV at the Tevatron and about 5~GeV at the LHC. The differential cross
sections decrease sharply with the increase of $q_T$, which indicates that most events
will happen in the relatively low $q_T$ region, where the resummation effects are
essential. For example, about 80 percent of events lie in the $q_T<20$~GeV region at the
Tevatron, while at the LHC the range is $q_T<40$~GeV.  Therefore, the precise theoretical
predictions for the distribution in the small $q_T$ region are very important, which can
help the selection of the the experimental cuts for better background rejection.

\section{Summary}\label{sec:summary}

In summary, we have calculated the next-to-leading order total cross section and
transverse momentum distribution of a single slepton in the MSSM without R-parity,
including all-order soft gluon resummation effects. For the total cross section, our
results are consistent with the results of Ref.~\cite{Nucl.Phys.B660.343}. We update
their numerical results with the updated version of the parton distribution functions and
find that the discrepancy between the different sets of PDFs has been decreased compared
with the previous version.  Also, we find that the factorization and renormalization
scale dependece in next-to-leading order is much smaller than the leading-order results.
For the transverse momentum distribution, we resummed the large logarithms at small
transverse momentum. Combined with the fixed order calculations, we give consistent
predictions for both small $q_T$ and large $q_T$. We also compared two approaches to the
non-perturbative parametrization and found that the results are slightly different at the
Tevatron and are in good agreement at the LHC.  Our result can be useful to the
simulation of the events and to the future collider experiments.

\begin{acknowledgments}
  We would like to thank Jianwei Qiu for very helpful discussions. This work was
  supported in part by the National Natural Science Foundation of China, under grant
  Nos.~10421003 and 10575001, and the Key Grant Project of Chinese Ministry of Education,
  under grant NO.~305001.
\end{acknowledgments}


\begin{thebibliography}{99}

\bibitem{hep-ph/0406039}
  R.~Barbier {\it et al.},
  arXiv:hep-ph/0406039.

\bibitem{hep-ph/0406029}
  M.~Chemtob,
  Prog.\ Part.\ Nucl.\ Phys.\  {\bf 54}, 71 (2005)
  [arXiv:hep-ph/0406029].

\bibitem{epem}
  S.~Dimopoulos and L.~J.~Hall,
  Phys.\ Lett.\ B {\bf 207}, 210 (1988).

  V.~D.~Barger, G.~F.~Giudice and T.~Han,
  Phys.\ Rev.\ D {\bf 40}, 2987 (1989).

  G.~F.~Giudice {\it et al.},
  arXiv:hep-ph/9602207.

  E.~Accomando {\it et al.}  [ECFA/DESY LC Physics Working Group],
  Phys.\ Rept.\  {\bf 299}, 1 (1998)
  [arXiv:hep-ph/9705442].

  J.~Erler, J.~L.~Feng and N.~Polonsky,
  Phys.\ Rev.\ Lett.\  {\bf 78}, 3063 (1997)
  [arXiv:hep-ph/9612397].

  J.~Kalinowski, R.~Ruckl, H.~Spiesberger and P.~M.~Zerwas,
  Phys.\ Lett.\ B {\bf 406}, 314 (1997)
  [arXiv:hep-ph/9703436].

\bibitem{Phys.Rev.D41.2099}
  S.~Dimopoulos, R.~Esmailzadeh, L.~J.~Hall and G.~D.~Starkman,
  Phys.\ Rev.\ D {\bf 41}, 2099 (1990).

\bibitem{sqhh}
  H.~K.~Dreiner and G.~G.~Ross,
  Nucl.\ Phys.\ B {\bf 365}, 597 (1991).

  B.~Allanach {\it et al.}  [R parity Working Group Collaboration],
  arXiv:hep-ph/9906224.

  H.~K.~Dreiner, P.~Richardson and M.~H.~Seymour,
  JHEP {\bf 0004}, 008 (2000)
  [arXiv:hep-ph/9912407].

  A.~Datta, J.~M.~Yang, B.~L.~Young and X.~Zhang,
  Phys.\ Rev.\ D {\bf 56}, 3107 (1997)
  [arXiv:hep-ph/9704257].

  J.~M.~Yang {\it et al.},
  arXiv:hep-ph/9802305.

  R.~J.~Oakes, K.~Whisnant, J.~M.~Yang, B.~L.~Young and X.~Zhang,
  Phys.\ Rev.\ D {\bf 57}, 534 (1998)
  [arXiv:hep-ph/9707477].

  E.~L.~Berger, B.~W.~Harris and Z.~Sullivan,
  Phys.\ Rev.\ Lett.\  {\bf 83}, 4472 (1999)
  [arXiv:hep-ph/9903549].

  T.~Plehn,
  Phys.\ Lett.\ B {\bf 488}, 359 (2000)
  [arXiv:hep-ph/0006182].

\bibitem{Nucl.Phys.B397.3}
  J.~Butterworth and H.~K.~Dreiner,
  Nucl.\ Phys.\ B {\bf 397}, 3 (1993)
  [arXiv:hep-ph/9211204].

\bibitem{slhh}
  G.~Moreau, M.~Chemtob, F.~Deliot, C.~Royon and E.~Perez,
  Phys.\ Lett.\ B {\bf 475}, 184 (2000)
  [arXiv:hep-ph/9910341].

  F.~Deliot, G.~Moreau and C.~Royon,
  Eur.\ Phys.\ J.\ C {\bf 19}, 155 (2001)
  [arXiv:hep-ph/0007288].

\bibitem{Phys.Rev.D63.055008}
  H.~K.~Dreiner, P.~Richardson and M.~H.~Seymour,
  Phys.\ Rev.\ D {\bf 63}, 055008 (2001)
  [arXiv:hep-ph/0007228].

\bibitem{Nucl.Phys.B660.343}
  D.~Choudhury, S.~Majhi and V.~Ravindran,
  Nucl.\ Phys.\ B {\bf 660}, 343 (2003)
  [arXiv:hep-ph/0207247].

\bibitem{Phys.Rep.58.269}
  Y.~L.~Dokshitzer, D.~Diakonov and S.~I.~Troian,
  Phys.\ Rept.\  {\bf 58}, 269 (1980).

\bibitem{Nucl.Phys.B154.427}
  G.~Parisi and R.~Petronzio,
  Nucl.\ Phys.\ B {\bf 154}, 427 (1979).

\bibitem{colsop}
  J.~C.~Collins and D.~E.~Soper,
  Nucl.\ Phys.\ B {\bf 193}, 381 (1981)
  [Erratum-ibid.\ B {\bf 213}, 545 (1983)].

  J.~C.~Collins and D.~E.~Soper,
  Nucl.\ Phys.\ B {\bf 197}, 446 (1982).

\bibitem{Nucl.Phys.B250.199}
  J.~C.~Collins, D.~E.~Soper and G.~Sterman,
  Nucl.\ Phys.\ B {\bf 250}, 199 (1985).

\bibitem{Phys.Rev.D67.073016}
  F.~Landry, R.~Brock, P.~M.~Nadolsky and C.~P.~Yuan,
  Phys.\ Rev.\ D {\bf 67}, 073016 (2003)
  [arXiv:hep-ph/0212159].

\bibitem{Phys.Rev.D63.114011}
  J.~w.~Qiu and X.~f.~Zhang,
  Phys.\ Rev.\ D {\bf 63}, 114011 (2001)
  [arXiv:hep-ph/0012348].


\end{thebibliography}
\end{document}